\documentclass[a4paper,10pt]{article}
\pdfoutput=1
\usepackage{amsmath}
\usepackage{amssymb}
\usepackage{mathtools}
\usepackage{graphicx}
\usepackage{xcolor}
\usepackage{footmisc}
\usepackage[normalem]{ulem}
\usepackage{float}
\usepackage{jcappub}
\usepackage{braket}
\usepackage{wrapfig}
\usepackage{lscape}
\usepackage{rotating}
\usepackage{epstopdf}
\usepackage{comment}
\usepackage[numbers, sort&compress]{natbib}
\usepackage{tikz, pgfplots}
\usetikzlibrary{positioning}
\usetikzlibrary{decorations.pathmorphing}
\usepackage[export]{adjustbox}[2011/08/13]

\tikzset{snake/.style={decorate, decoration=snake}}

\def\normOrd#1{\mathop{:}\nolimits\!#1\!\mathop{:}\nolimits}

\newcommand{\be}{\begin{equation}}
	\newcommand{\ee}{\end{equation}}

\begin{document}
	\title{Cosmological perturbation theory for large scale structure in phase space}
	\author{Caio Nascimento and Marilena Loverde}
	\affiliation{Department of Physics, University of Washington, Seattle, WA, USA}
	\emailAdd{caiobsn@uw.edu}

\abstract{
We develop a framework for Large Scale Structure (LSS) perturbation theory, that solves the Vlasov-Poisson system of equations for the distribution function in full phase space. This approach relaxes the usual \textit{apriori} assumption of negligible velocity dispersion underlying the Standard Perturbation Theory (SPT). We apply the new method to rederive the usual SPT kernels up to third order in the perturbative expansion. We also show that a counterterm, identical to the one introduced by standard Effective Field Theory (EFT) methods, naturally arises within our framework. We finish by making a precise connection to EFT techniques, which reveals the necessity of the EFTofLSS to self-consistently model the long-wavelength fluid, and illustrates the importance of having theoretical control over short distance fluctuations.}

\maketitle

\section{Introduction}
\label{sec:int}

Perturbation theory methods for the evolution of large-scale structure in the universe play a central role in our understanding of gravitational instability in the nonlinear regime \cite{Makino:1991rp, Jain:1993jh, Goroff:1986ep, Zeldovich:1969sb, Bernardeau:2001qr, Scoccimarro:1995if}. In one hand, they provide flexibility to go beyond standard scenarios \cite{Bottaro:2023wkd, Piga:2022mge, Lewandowski:2019txi, Crisostomi:2019vhj, Bottaro:2024pcb} and valuable intuition on the nonlinear gravitational evolution. They also do not require expensive computational resources, contrary to simulations \cite{Schneider:2015yka}. On the other hand, perturbation theory methods have a limited range of applicability as N-body simulations are required to accurately model the dynamics on sufficiently small scales \cite{Carlson:2009it}. 

The collective dynamics of N-body particles in an expanding background, interacting solely via gravity, is encoded in the collisionless Boltzmann (or Vlasov) equation in phase space coupled to the Poisson equation, the Vlasov-Poisson system. The standard perturbation theory (SPT) approach is then based on a truncation of the Boltzmann hierarchy in its first two moments, which corresponds to the assumption of an ideal pressureless fluid \cite{Bernardeau:2001qr}.  An alternative but equivalent approach to model the dynamics of late time cosmological fluctuations perturbatively is Lagrangian Perturbation Theory (LPT), where one solves for the displacement field connecting the initial (Lagrangian) and final (Eulerian) particle positions \cite{Buchert:1993ud, Catelan:1994ze, Bouchet:1994xp} \footnote{It can be shown that LPT matches SPT order by order in the perturbative expansion (e.g. \cite{Sugiyama:2013mpa}). See \cite{Vlah:2015zda} for a thorough comparison between these two approaches.}.

Over the years many improvements to SPT and LPT have been proposed, and some of them are now a central piece of the theory modeling involved in the analysis of real data \cite{Euclid:2023tog,Chudaykin:2020aoj, Ivanov:2019pdj,Philcox:2022frc, Taule:2023izt, Beutler:2011hx, Blake:2011rj, Chudaykin:2022nru, Taule:2024bot, Aviles:2024zlw, Colas:2019ret, Simon:2022csv, DAmico:2022osl, DESI:2024jis, Chen:2024vvk, Chen:2022jzq, Moretti:2023drg}.  Such improvements can be broadly divided into two classes. The first corresponds to the set of tools that do not attempt to describe an imperfect fluid, but are rather based on a reorganization of the perturbative expansion and resummations of certain classes of diagrams to all orders in perturbation theory. Examples of such methods are Renormalized Perturbartion Theory (RPT) \cite{Crocce:2005xy, Crocce:2007dt, Taruya:2007xy, Bernardeau:2008fa, Taruya:2012ut} and Infrared Resummation (IR) schemes \cite{Ivanov:2019dlz, Sugiyama:2024eye, Baldauf:2015xfa, Lewandowski:2018ywf, Blas:2016sfa, Pietroni:2008jx, Chen:2024pyp, Chen:2020zjt, Matsubara:2007wj}. 

In this work our goal will be to shed light on the second class of improvements to SPT, i.e.  those that accommodate the inevitable deviations from a pressureless fluid that arise during nonlinear structure formation. In this class the leading framework is the Effective Field Theory of Large Scale Structure (EFTofLSS) \cite{Baumann:2010tm, Carrasco:2012cv, Carroll:2013oxa, Pietroni:2011iz, Carrasco:2013mua, Porto:2013qua, Vlah:2015sea, Zaldarriaga:2015jrj}, which also addresses the issue of sensitivity to uncontrolled short distance physics \footnote{As can be seen from the fact that modes with arbitrarily high frequencies are running on the loops in SPT.}, and has been very successful in pushing the regime of validity of perturbation theory methods towards smaller scales \cite{Baldauf:2016sjb, Braganca:2023pcp, Nishimichi:2020tvu}. It does so, however, at the cost of introducing new nuisance parameters to the theoretical model.

We will be focusing on the simplest case of perturbation theory to one-loop, describing the two-point clustering of matter in real space (as opposed to redshift space). In that case only a single new parameter is needed, the effective sound speed, which can be determined either by observations or through matching to N-body simulations. Additional free parameters become necessary when considering higher order terms in the perturbative expansion \cite{Konstandin:2019bay}, when modeling biased tracers \cite{Desjacques:2016bnm} and for higher point correlation functions \cite{Baldauf:2014qfa, Baldauf:2021zlt, Bertolini:2016bmt} as well. 

On the other hand, gravity-only N-body simulations have no free parameters, and we are entering a new era where efficient emulators are available to interpolate the predictions from simulations in broad regions of parameter (and even theory) space  \cite{Euclid:2020rfv, Angulo:2020vky, Heitmann:2013bra, Nishimichi:2018etk, DeRose:2018xdj, Wibking:2017slg, Winther:2019mus, Brando:2022gvg, Mauland:2023pjt, Jamieson:2024fsp, Jamieson:2022lqc}. This motivates the search for a theoretical framework which does not introduce new nuisance parameters. The starting point of SPT is the assumption of vanishing velocity dispersion (which corresponds to an ideal fluid), and this leaves open the possibility that one can account for the dissipative terms in a fully perturbative framework that truncates the Boltzmann hierarchy at a higher than second moment \cite{Garny:2022tlk, Garny:2022kbk, Garny:2025ovs, Erschfeld:2018zqg, Dominguez:2000dt,mcdonald2011generate, Kitaura:2023mfc, Matarrese:2007wc, McDonald:2017ths, Aviles:2015osc, Bharadwaj:1995px}.

In what follows we will pursue the question of how far one can go with old-fashioned cosmological perturbation theory methods, without using Effective Field Theory (EFT) ingredients, in terms of accurately predicting observables of interest. To accomplish this we will develop a framework to perturbatively solve the Vlasov-Poisson system of equations directly in phase space, expanding upon ideas first introduced in \cite{Valageas:2001df}. This framework circumvents the need to artificially truncate the Boltzmann hierarchy and hence relaxes the usual \textit{apriori} assumptions of negligible vorticity and velocity dispersion (see \cite{tassev2011helmholtz, Valageas:2003gm} for other previous approaches to nonlinear structure formation in phase space). 

We will first use this framework to rederive the familiar SPT kernels. This exercise underscores the fact that a vanishing vorticity and velocity dispersion should be seen as a consequence of the perturbative expansion, rather than an assumption (in accordance with the results obtained in \cite{Valageas:2001df}). This is consistent with the expectation that these effects are of intrinsically nonperturbative nature. An important outcome of our formalism will be the fact that nonlinearities backreact into the background distribution function. That is, even if our starting point is a background distribution function consistent with the assumption of cold dark matter (proportional to a Dirac delta function at zero momentum), gravitational nonlinearities will introduce some additional time-dependent contributions.  In fact, we will see that the Boltzmann equation can be rephrased as a coupled set of equations, one for the background distribution function and another for its fluctuations. 

This suggests a natural procedure to improve on SPT: To use the (\textit{apriori} unknown) fully nonlinear background distribution function as a source term to solve for the fluctuation in the distribution function perturbatively. This effectively enables one to account for a nonzero average velocity dispersion, and introduces an additional contribution to the one-loop power spectrum with the exact same form as the effective sound speed counterterm in the EFTofLSS. While this reveals the possibility to arrive at the right ingredients from a purely old-fashioned perturbation theory approach, we will show that the framework is necessarily incomplete for reasons that are related to the lack of theoretical control over short distance fluctuations. This problem can be cured by standard EFT methods, which emerge as a necessary ingredient to self-consistently model a nonzero velocity dispersion.      

We will be focusing on the minimal scenario of a $\Lambda$CDM universe, particularly its late time dynamics (redshifts $z \lesssim  100$) on subhorizon scales $k \gg aH$ (where $a(t)$ is the cosmological scale factor, $H=d\log a/dt$ is the Hubble expansion rate and $t$ is cosmic time) \footnote{We do however wish to consider time scales that go beyond a Hubble time $t \gtrsim t_{H} \sim 1/H$.}. All numerical calculations assume a fiducial $\Lambda$CDM cosmology with $\Omega_{\textrm{m},0} =0.3$, $\Omega_{\Lambda,0}=0.7$ and $h=0.7$, where $H_{0}=100h\textrm{km/s}/\textrm{Mpc}$ is the Hubble expansion rate today.

We structure this paper as follows: In Sec.\ref{sec:spt} we review the framework underlying Standard Perturbation Theory (SPT), to establish notation and for later comparison with our new perturbation theory scheme based directly on the Vlasov-Poisson system of equations in phase space, developed in Sec.\ref{sec:ptbe}. In Sec.\ref{sec:eftoflss} we first show how an EFT-like counterterm naturally emerges within our framework, and then proceed to make the connection to EFT methods more precise, which underscores the necessity of the EFTofLSS to self-consistently model a nonzero velocity dispersion. Our main results are summarized in Sec.\ref{sec:concl}. Additionally, Appendices \ref{sec:app1}, \ref{sec:app2} and \ref{sec:app3} derive important equations which are used in the main text. Appendices \ref{sec:app4} and \ref{sec:app5} contain explicit formulas for the one-loop power spectrum and tree-level bispectrum in SPT, to aid the reader with numerical calculations involving the full time dependencies of perturbation theory kernels in $\Lambda$CDM.

\section{Standard Perturbation Theory}
\label{sec:spt}

The collective behavior of particles interacting only gravitationally in an expanding universe is governed by the collisionless Boltzmann, or Vlasov, equation
\begin{equation}
\label{eq:vlasov}
	\frac{\partial f}{\partial \eta} + \frac{d\vec{x}}{d\eta} \cdot \frac{\partial f}{\partial \vec{x}} + \frac{d\vec{q}}{d\eta} \cdot \frac{\partial f}{\partial \vec{q}} = 0 \, ,
\end{equation}
for the phase space distribution function $f(\eta, \vec{x},\vec{q})$. We work with the superconformal time defined by $d\eta = dt/a^2(t)$.  Additionally, $\vec{x}$ are comoving coordinates and $\vec{q}$ is the comoving momentum such that $(d\vec{x}/d\eta) = a^2 (d\vec{x}/dt) = (\vec{q}/m)$, with $m$ the particle mass. We then have $(d\vec{q}/d\eta) = -ma^2 \vec{\nabla} \phi$, with $\phi(\eta,\vec{x})$ the gravitational potential. Eq.(\ref{eq:vlasov}) now reads,
\begin{equation}
\label{eq:vlasov2}
		\frac{\partial f}{\partial \eta} + \frac{\vec{q}}{m} \cdot \frac{\partial f}{\partial \vec{x}} -m a^2(\eta) \frac{\partial \phi}{\partial \vec{x}} \cdot \frac{\partial f}{\partial \vec{q}} = 0 \, .
\end{equation}
Next we take moments of Eq.(\ref{eq:vlasov2}) to arrive at fluid equations, following standard procedure \cite{Bernardeau:2001qr}. First define the energy density
\begin{equation}
\label{eq:energydensity}
	\rho(\eta, \vec{x}) = ma^{-3}(\eta) \int \frac{d^3 \vec{q}}{(2\pi)^{3}} f(\eta,\vec{x},\vec{q}) \,.
\end{equation}
The derivative of Eq.(\ref{eq:energydensity}) with respect to $\eta$, which we here denote by a prime, yields using Eq.(\ref{eq:vlasov2})
\begin{equation}
\label{eq:continuity}
	\rho' + 3\mathcal{H} \rho +a \vec{\nabla} \cdot \vec{\Pi} = 0 \, ,
\end{equation}
where $\mathcal{H} = d\log a/d\eta = a^2H$ and
\begin{equation}
\label{eq:momentum}
	\vec{\Pi}(\eta,\vec{x}) = ma^{-4}(\eta) \int \frac{d^3 \vec{q}}{(2\pi)^{3}} \, \vec{q} \, f(\eta,\vec{x},\vec{q}) \,, 
\end{equation}
is the fluid momentum. We once again take a derivative of Eq.(\ref{eq:momentum}) with respect to superconformal time, and use Eq.(\ref{eq:vlasov2}) to arrive at
\begin{equation}
\label{eq:euler} 
\Pi_{i}' + 4\mathcal{H} \Pi_{i} +2a \partial^{j} K_{ij} +a\rho \partial_{i} \phi = 0 \,,
\end{equation}
when written in terms of a kinetic energy density tensor
\begin{equation}
\label{eq:stress}
	K_{ij}(\eta,\vec{x}) =  \frac{1}{2m} a^{-5}(\eta) \int \frac{d^3 \vec{q}}{(2\pi)^{3}} \, q_{i}q_{j} \, f(\eta,\vec{x},\vec{q}) \,.
\end{equation}
Eqs.(\ref{eq:continuity}) and (\ref{eq:euler}) are called the first and second moments of the Boltzmann equation, corresponding to the continuity and Euler equations respectively. Note that these moments do not depend explicitly on the particle mass $m$, which in fact can be absorbed into suitable redefinitions of the comoving momentum $\vec{q}$ and distribution function $f$ as follows: $\vec{q} \to m \vec{q}$ and $f \to f/m^{4}$. For this reason we simply set $m=1$ moving forward.

Both Eqs.(\ref{eq:continuity}) and (\ref{eq:euler}) pick-up corrections that are sizable on horizon scales and for large thermal velocities that can appear, for example, in warm dark matter models. For instance, note that the energy density in Eq.(\ref{eq:energydensity}) is in reality a mass density since it only includes the rest mass contribution to the particle proper energy. Such corrections are negligible for cold dark matter and at sub-horizon scales. 

In principle we can continue by taking the derivative of Eq.(\ref{eq:stress}) with respect to superconformal time, to arrive at an equation of motion for the second moment, involving the third moment of the Boltzmann equation. Repeating this procedure indefinitely generates coupled equations of motion for even higher moments, the so-called Boltzmann hierarchy. Instead, the starting point of SPT is to truncate the resulting Boltzmann hierarchy at its second moment. To see how this works let us first introduce the field velocity $\vec{v}(\eta,\vec{x})$ as follows, 
\begin{equation}
\label{eq:velocity}
	\vec{\Pi}(\eta,\vec{x}) = \rho(\eta,\vec{x}) \vec{v}(\eta,\vec{x})\,, 
\end{equation}
in terms of which the Euler Eq.(\ref{eq:euler}) becomes
\begin{equation}
\label{eq:euler2}
	v_{i}' + \mathcal{H} v_{i} +a v_{j} \partial^{j} v_{i} + a \partial_{i} \phi + \frac{a}{\rho} \partial^{j} \tau_{ij} = 0 \,,
\end{equation}
after substituting Eq.(\ref{eq:velocity}) into Eq.(\ref{eq:euler}) and dividing by the energy density. This involves the stress tensor
\begin{equation}
\label{eq:stresstensor}
\begin{split}
	 \tau_{ij}&(\eta,\vec{x})  = 2K_{ij}(\eta,\vec{x}) - \rho(\eta,\vec{x}) v_{i}(\eta,\vec{x}) v_{j}(\eta,\vec{x}) \\ & = a^{-5}(\eta) \int \frac{d^3 \vec{q}}{(2\pi)^{3}} \, \left[q_{i}-a(\eta)v_{i}(\eta,\vec{x})\right]\left[q_{j}-a(\eta)v_{j}(\eta,\vec{x})\right] f(\eta,\vec{x},\vec{q}) \,.
\end{split}
\end{equation}
Note from its definition in the second line of Eq.(\ref{eq:stresstensor}) that the stress tensor is sourced by velocity dispersion at a fixed comoving position (dispersion with respect to averaging over momentum at a given point in configuration space), and hence can only be nonvanishing at the onset of shell-crossing when particle trajectories intersect.

In order to turn Eqs.(\ref{eq:continuity}) and (\ref{eq:euler2}) into a closed system, we first need to introduce the Poisson equation satisfied by the gravitational potential:
\begin{equation}
\label{eq:poisson}
	\nabla^2 \phi = 4\pi Ga^2 (\rho-\bar{\rho}) \,,
\end{equation}
where only the fluctuations around the average density $\bar{\rho}(\eta) = \langle \rho(\eta,\vec{x}) \rangle \propto a^{-3}(\eta)$ contribute to the gravitational potential. Eq.(\ref{eq:poisson}) also picks-up corrections on large scales and for large thermal velocities, but they are negligible for cold dark matter on sub-horizon scales.  

One next proceeds with the standard assumption of a negligible stress tensor: $\tau_{ij} \approx 0$. Under this assumption the velocity field is fully specified by its divergence, i.e., the vorticity degrees of freedom are negligible. To see why that is, define $w(\eta,\vec{x})=\vec{\nabla} \times \vec{v} (\eta,\vec{x})$, and take the curl of Eq.(\ref{eq:euler2}) to obtain
\begin{equation}
	\label{eq:vortex}
	w_{i}' + \mathcal{H} w_{i} -a [\vec{\nabla} \times (\vec{v} \times \vec{w})]_{i} = - a \epsilon_{i}^{jk} \partial_{j} \left(\frac{1}{\rho} \partial^{l} \tau_{kl}\right)  \,.
\end{equation}
The source term in the right-hand side of Eq.(\ref{eq:vortex}) vanishes when $\tau_{ij} \approx 0$, such that vorticity can be neglected if it is not present in the initial conditions. This latter assumption is justified since the vorticity decays with the expansion of the universe in linear perturbation theory \cite{Bernardeau:2001qr}. We can even relax the assumption of $\tau_{ij} \approx 0$, and consider a diagonal stress tensor of the form $\tau_{ij} = p \delta_{i,j}$, where $p=p(\eta,\vec{x})$ is the pressure. In this case
\begin{equation}
	\label{eq:vortex_source}
 - \epsilon_{i}^{jk} \partial_{j} \left(\frac{1}{\rho} \partial^{l} \tau_{kl}\right) = \left(\frac{\vec{\nabla} \rho}{\rho^{2}} \times \vec{\nabla} p\right)_{i} \,,
\end{equation}
which vanishes whenever the pressure is an arbitrary function of the density, $p=p(\rho)$, as in an adiabatic fluid. We then once again reach the conclusion that vorticity can be neglected \cite{Pueblas:2008uv}.   

Once both the stress tensor and vorticity are neglected, the continuity and Euler Eqs.~(\ref{eq:continuity}) and (\ref{eq:euler2}) can be solved perturbatively in fluctuations around the homogeneous background, see \cite{Bernardeau:2001qr}. SPT accurately describes the clustering of matter in single-stream regions (often denoted by voids) where particle trajectories do not intersect. In such regions the stress tensor vanishes. However, at sufficiently small scales the gravitational evolution becomes strongly coupled, leading to shell-crossing and the subsequent formation of bound structures via gravitational collapse where particle trajectories do intersect \cite{Zeldovich:1969sb, doi:10.1080/03091928208209001}. This is signaled by the emergence of a nonzero stress tensor. 

While on large scales isotropy is a good approximate symmetry of the perturbations, the local process of gravitational collapse is anisotropic and happens at different rates along different axes, as determined by the engenvectors of the tidal tensor $\partial_{i}\partial_{j} \phi$ \footnote{The recent paper \cite{Musso:2024roa} argues for an energy shear tensor criteria.}. The gravitational collapse then proceeds in a hierarchical triaxial way: First into cosmic sheets, followed by cosmic filaments until the remaining axis finally collapses and dark matter halos form. The outcome is an intricate cosmic web where dark matter halos can be found within filaments which themselves can be found within sheets \cite{Jaber:2023rjx, Aragon-Calvo:2023oxm, Libeskind:2017tun, Cautun:2014fwa, Kugel:2024zxq}. 

Cosmic sheets, filaments and halos all correspond to multi-stream regions where the stress tensor does not vanish \cite{Buehlmann:2018qmm}, and hence SPT breaks down. Note that although $\tau_{ij}=0$ is an \textit{apriori} assumption of the standard perturbative framework, we expect the emergence of a nonzero stress to be an intrinsically nonperturbative phenomena in nonlinear gravitational evolution \footnote{A nonzero vorticity is also generated by nonperturbative effects \cite{Jelic-Cizmek:2018gdp, Pichon:1999tk}, and we can think of it as contributing to the effective stress tensor.}. Indeed, in the next section we will see that the Vlasov-Poisson system of Eqs.(\ref{eq:vlasov2}) and (\ref{eq:poisson}) can be solved perturbatively in full phase space, allowing us to relax the \textit{apriori} assumption of a vanishing stress tensor.

\section{Cosmological perturbations in phase space}
\label{sec:ptbe}

In this section we will develop a framework to directly solve the Vlasov-Poisson system of equations perturbatively in full phase space. It will then become clear that a negligible stress is a consequence of the perturbative expansion, rather than an assumption. We will also show that small scale nonlinearities backreact into the background distribution function, which enables one to naturally account for a nonzero average velocity dispersion by introducing the unknown fully nonlinear background distribution function into the formalism. This adds a new term to the nonlinear power spectrum which has the exact same form as the effective sound speed counterterm in the EFTofLSS.

\subsection{An iterative solution to Vlasov-Poisson}
\label{sec:iter}

We first present a derivation of SPT based on a perturbative solution to the Vlasov-Poisson system of equations in full phase space, extending upon the work of \cite{Valageas:2001df}. Let us repeat here for convenience the collisionless Boltzmann (or Vlasov) Eq.(\ref{eq:vlasov2}) 
\begin{equation}
	\label{eq:vlasov3}
	\frac{\partial f}{\partial \eta} + \vec{q} \cdot \frac{\partial f}{\partial \vec{x}} =a^2(\eta) \frac{\partial \phi}{\partial \vec{x}} \cdot \frac{\partial f}{\partial \vec{q}} \,,
\end{equation}
where we set $m=1$\footnote{A justification for this choice can be found in Sec.~\ref{sec:spt}, in the discussion below Eq.~(\ref{eq:stress}).}, and moved the nonlinear term to the right-hand side for future convenience. At first we will remain agnostic about what is sourcing the gravitational potential $\phi(\eta,\vec{x})$, so we will delay writing down the Poisson Eq.(\ref{eq:poisson}). 

Next we split Eq.~(\ref{eq:vlasov3}) into a coupled set of equations, one for the background distribution function $\bar{f}(\eta,q) = \langle f(\eta,\vec{x},\vec{q}) \rangle$ defined by an ensemble average \footnote{This can also be thought of as a volume average. A precise operational definition will be given later in this section.}, and another for its fluctuations $\delta f=f-\bar{f}$. This procedure is not strictly necessary as Eq.~(\ref{eq:vlasov3}) can be solved perturbatively as is, but it will prove useful for later developments. The ensemble average of Eq.~(\ref{eq:vlasov3}) reads
\begin{equation}
\label{eq:vlasovavg}
	\frac{\partial \bar{f}}{\partial \eta} = a^2(\eta) \Big\langle \frac{\partial \phi}{\partial \vec{x}} \cdot \frac{\partial f}{\partial \vec{q}} \Big\rangle \,,
\end{equation}
where we used the fact that $\bar{f}(\eta,q)$ is position independent to drop the term proportional to its spatial gradient \footnote{Or thinking in terms of a volume average, that term becomes a total derivative and hence leads to a surface contribution, which we assume vanishes at spatial infinity with suitable boundary conditions.}. Note that the right-hand side of Eq.(\ref{eq:vlasovavg}) includes the ensemble average of a term quadratic in fluctuations and is hence nonvanishing. For this reason small scale nonlinearities will backreact into the background distribution function. Subtracting Eq.(\ref{eq:vlasovavg}) from Eq.(\ref{eq:vlasov3}) produces the equation for the fluctuation to the distribution function $\delta f(\eta,\vec{x},\vec{q})$:
\begin{equation}
\label{eq:vlasovfluc}
	\frac{\partial \delta f}{\partial \eta} + \vec{q} \cdot \frac{\partial \delta f}{\partial \vec{x}} =a^2(\eta) \normOrd{ \frac{\partial \phi}{\partial \vec{x}} \cdot \frac{\partial f}{\partial \vec{q}}} \,,
\end{equation}
where we introduced the normal ordering symbol as subtracting ensemble averages, that is,
\begin{equation}
\label{eq:normalordering}
	\normOrd{ \frac{\partial \phi}{\partial \vec{x}} \cdot \frac{\partial f}{\partial \vec{q}}} =  \frac{\partial \phi}{\partial \vec{x}} \cdot \frac{\partial f}{\partial \vec{q}} -  \Big\langle \frac{\partial \phi}{\partial \vec{x}} \cdot \frac{\partial f}{\partial \vec{q}} \Big\rangle \,.
\end{equation}
Since $f=\bar{f} + \delta f$ appears on the right-hand side of Eqs.~(\ref{eq:vlasovavg}) and (\ref{eq:vlasovfluc}) these are a coupled set of equations, which we will now write in their integral forms. For Eq.(\ref{eq:vlasovavg}) this is straightforward and follows from an integration over superconformal time
\begin{equation}
\label{eq:vlasovavg2}
	\bar{f}(\eta,q) = \bar{f}^{(0)}(q) + \int_{0}^{\eta} d\eta' a^2(\eta') \Big\langle \frac{\partial \phi}{\partial \vec{x}} \cdot \frac{\partial f}{\partial \vec{q}} \Big\rangle \Big|_{\eta'} \,,
\end{equation}
where $\bar{f}^{(0)}(q)$ is the background distribution function before picking up nonlinear corrections. For a cold dark matter species, $ \bar{f}^{(0)}(q) \propto \delta^{(3)}(\vec{q})$.

In Fourier space \footnote{Let $\vec{\nabla} \to i\vec{k}$.} Eq.(\ref{eq:vlasovfluc}) becomes a first order ODE and it is then a straightforward exercise to rephrase it as an integral equation:
\begin{equation}
\label{eq:integralform}
	\delta f(\eta,\vec{k},\vec{q}) = \int_{0}^{\eta} d\eta' a^2(\eta') e^{-i\vec{k}\cdot \vec{q}(\eta-\eta')} \left[ \normOrd{ \frac{\partial \phi}{\partial \vec{x}} \cdot \frac{\partial f}{\partial \vec{q}}} \right]\Bigg|_{\eta',\vec{k}} \,,
\end{equation}
where \footnote{Note that we also set $\delta f(\eta=0,\vec{k},\vec{q})=0$. This is justified here since the initial conditions only play a role at horizon scales while at the subhorizon scales of interest the source term completely dominates.}
\begin{equation}
\label{eq:convo}
	\left[ \normOrd{ \frac{\partial \phi}{\partial \vec{x}} \cdot \frac{\partial f}{\partial \vec{q}}} \right]\Bigg|_{\eta',\vec{k}} =  \int \frac{d^3 \vec{k}_1}{(2\pi)^3} \frac{d^3 \vec{k}_2}{(2\pi)^3} (2\pi)^3 \delta^{(3)}(\vec{k}-\vec{k}_1 -\vec{k}_2) \normOrd{ \phi(\eta',\vec{k}_1) i\vec{k}_1 \cdot \frac{\partial f}{\partial \vec{q}}\Big|_{\eta',\vec{k}_2,\vec{q}}} \,, \small
\end{equation}
denotes a convolution. These are the ingredients we need to start solving this coupled set of Boltzmann equations in phase space. 

We are now ready to consider an iterative solution to Eqs.(\ref{eq:vlasovavg2}) and (\ref{eq:integralform}) in the form of
\begin{equation}
\label{eq:iterative}
\begin{split}
	& \bar{f}(\eta,q) = 	\bar{f}^{(\textrm{0th})}(\eta,q) + \bar{f}^{(\textrm{1st})}(\eta,q) + \bar{f}^{(\textrm{2nd})}(\eta,q) + \cdots \\ & \delta f(\eta,\vec{k},\vec{q}) = 	\delta f^{(\textrm{0th})}(\eta,\vec{k},\vec{q}) + \delta f^{(\textrm{1st})}(\eta,\vec{k},\vec{q}) + \delta f^{(\textrm{2nd})}(\eta,\vec{k},\vec{q}) + \cdots \,,
\end{split}
\end{equation}
for a given external gravitational potential $\phi(\eta,\vec{k})$. This can be represented diagrammatically using circles connected by a horizontal line, with the number of circles denoting the order in the iterative expansion. Additionally, we use a solid horizontal line to denote a term in the iterative expansion for the fluctuation in the distribution function, and a dashed horizontal line is used for the background distribution function. 

\begin{figure}
\centering
\begin{tikzpicture} 
	\draw	(0,0) -- (0.5,0) (0.75,0) circle(0.25) (1,0) -- (1.5,0) (1.75,0) circle(0.25) (2,0) -- (2.5,0) (2.75,0) circle(0.25) (3,0) -- (3.5,0) ;
\end{tikzpicture}
\caption{Diagram for the third order term in the iterative expansion for the fluctuation in the distribution function.}
\label{fig:diagram2}
\end{figure}
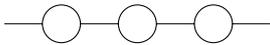

For example, the zeroth order term for the background distribution function, $\bar{f}^{(\textrm{0th})}(\eta,q)$,  corresponds to the diagram shown in Fig.\ref{fig:diagram1}. On the other hand, the third order term in the iterative expansion for the distribution function fluctuation, $\delta f^{(\textrm{3rd})}(\eta,\vec{k},\vec{q})$, is represented by the diagram drawn in Fig.\ref{fig:diagram2}.

\begin{figure}
\centering
\begin{tikzpicture} 
 \draw [dashed]	(0,0) -- (2,0);
\end{tikzpicture}
\caption{Diagram for the zeroth order term in the iterative expansion for the background distribution function. }
\label{fig:diagram1}
\end{figure}

An iterative solution of this form was considered previously in \cite{Senatore:2017hyk} (without the background-fluctuation split) in the context of massive neutrinos, and can be interpreted as a reconstruction of particle trajectories in powers of the gradient of the gravitational potential in units of the Hubble scale, i.e., $\sim \nabla^{2}\phi/H^2 $ is the effective expansion parameter (we will further comment on this at the end of this subsection). To zeroth order we simply have,
\begin{equation}
\label{eq:zeroth}
\begin{split}
	 & \bar{f}^{(\textrm{0th})}(\eta,q) = \bar{f}^{(0)}(q) \\ & \delta f^{(\textrm{0th})}(\eta, \vec{k}, \vec{q}) = 0 \,,
\end{split} 
\end{equation}
corresponding to the unperturbed trajectory. We now substitute this zeroth order solution into the right-hand side of Eqs.(\ref{eq:vlasovavg2}) and (\ref{eq:integralform}) to obtain the first order solutions. To evaluate the ensemble averages all one needs to know are the statistical properties of the external gravitational potential, which we take as a given. We arrive at,
\begin{equation}
\label{eq:first}
\begin{split}
	& \bar{f}^{(\textrm{1st})}(\eta, q) = 0 \,, \\ & \delta f^{(\textrm{1st})}(\eta, \vec{k}, \vec{q}) = i\vec{k} \cdot \frac{\partial \bar{f}^{(0)}}{\partial \vec{q}} \int_{0}^{\eta} d\eta' a^2(\eta') e^{-i\vec{k}\cdot \vec{q}(\eta-\eta')} \phi(\eta',\vec{k}) \,,
\end{split}
\end{equation}
where we assume $\langle \phi(\eta,\vec{k}) \rangle =0$. We can now repeat this procedure and substitute Eq.(\ref{eq:first}) into the right-hand side of Eqs.(\ref{eq:vlasovavg2}) and (\ref{eq:integralform}). Let us first stop for a moment to introduce notation which will be used throughout the manuscript,
\begin{equation}
\label{eq:notation}	
	\int \frac{d^3 \vec{k}_1}{(2\pi)^3} \cdots \frac{d^3 \vec{k}_N}{(2\pi)^3} (2\pi)^{3} \delta^{(3)}(\vec{k}-\vec{k}_1-\cdots -\vec{k}_N) \equiv \int_{\{\vec{k}_1, \cdots, \vec{k}_N\}}^{\vec{k}}\,,
\end{equation}
and in terms of which the second order solutions are
\begin{equation}
	\label{eq:second}
	\begin{split}
		& \bar{f}^{(\textrm{2nd})}(\eta, q) = \int_{0}^{\eta} d\eta' a^2(\eta') \int_{\{\vec{k}_1, \vec{k}_2\}}^{\vec{k}} \langle \phi(\eta',\vec{k}_1) i\vec{k}_1 \cdot \frac{\partial \delta f^{(\textrm{1st})}}{\partial \vec{q}}\Big|_{\eta',\vec{k}_2,\vec{q}} \, \rangle  \\ & \delta f^{(\textrm{2nd})}(\eta, \vec{k}, \vec{q}) = \int_{0}^{\eta} d\eta' a^2(\eta') e^{-i\vec{k}\cdot \vec{q}(\eta-\eta')} \int_{\{\vec{k}_1, \vec{k}_2\}}^{\vec{k}} \normOrd{\phi(\eta',\vec{k}_1) i\vec{k}_1 \cdot \frac{\partial \delta f^{(\textrm{1st})}}{\partial \vec{q}}\Big|_{\eta',\vec{k}_2,\vec{q}}} \,,
	\end{split}
\end{equation}
after plugging in the second line of Eq.(\ref{eq:first}) into Eq.(\ref{eq:second}). Here we do want to go up to third order since this is required for a one-loop calculation of the power spectrum. For that we substitute Eq.(\ref{eq:second}) into the right-hand side of Eqs.(\ref{eq:vlasovavg2}) and (\ref{eq:integralform}), which involves some additional terms when compared to the second order solution because $f^{(\textrm{2nd})} = \bar{f}^{(\textrm{2nd})} + \delta f^{(\textrm{2nd})}$ is the quantity that appears as a source, and now $\bar{f}^{(\textrm{2nd})} \neq 0$ (as opposed to  $\bar{f}^{(\textrm{1st})}=  0$ so that the background term does not contribute at second order). We arrive at,
\begin{equation}
	\label{eq:third1}
	\begin{split}
		& \bar{f}^{(\textrm{3rd})}(\eta, q) = \int_{0}^{\eta} d\eta' a^2(\eta') \int_{\{\vec{k}_1, \vec{k}_2\}}^{\vec{k}} \langle \phi(\eta',\vec{k}_1) i\vec{k}_1 \cdot \frac{\partial \bar{f}^{(\textrm{2nd})}}{\partial \vec{q}}\Big|_{\eta',\vec{k}_2,\vec{q}} \, \rangle  \\ & + \int_{0}^{\eta} d\eta' a^2(\eta') \int_{\{\vec{k}_1, \vec{k}_2\}}^{\vec{k}} \langle \phi(\eta',\vec{k}_1) i\vec{k}_1 \cdot \frac{\partial \delta f^{(\textrm{2nd})}}{\partial \vec{q}}\Big|_{\eta',\vec{k}_2,\vec{q}} \, \rangle \,,
	\end{split}
\end{equation}
and
\begin{equation}
	\label{eq:third2}
	\begin{split}
		& \delta f^{(\textrm{3rd})}(\eta, \vec{k}, \vec{q}) = \int_{0}^{\eta} d\eta' a^2(\eta') e^{-i\vec{k}\cdot \vec{q}(\eta-\eta')} \int_{\{\vec{k}_1, \vec{k}_2\}}^{\vec{k}} \normOrd{\phi(\eta',\vec{k}_1) i\vec{k}_1 \cdot \frac{\partial \bar{f}^{(\textrm{2nd})}}{\partial \vec{q}}\Big|_{\eta',\vec{k}_2,\vec{q}}}   \\ & + \int_{0}^{\eta} d\eta' a^2(\eta') e^{-i\vec{k}\cdot \vec{q}(\eta-\eta')} \int_{\{\vec{k}_1, \vec{k}_2\}}^{\vec{k}} \normOrd{\phi(\eta',\vec{k}_1) i\vec{k}_1 \cdot \frac{\partial \delta f^{(\textrm{2nd})}}{\partial \vec{q}}\Big|_{\eta',\vec{k}_2,\vec{q}}}  \,.
	\end{split}
\end{equation}
Note that it is only at third order in this expansion that the background distribution function backreacts into the fluctuations (this will be important later). The recursion relations found in Eqs.(\ref{eq:third1}) and (\ref{eq:third2}) can be straightforwardly generalized to higher orders in the iterative expansion, and for a known external gravitational potential this is the full story (and this is the extent to which the iterative solution was considered in \cite{Senatore:2017hyk}, since one can safely ignore the backreaction of neutrino fluctuations to the total gravitational potential which is dominated by cold dark matter).

However, in practice we know that the Poisson equation relates the gravitational potential to the density field, which is itself obtained from the distribution function via a momentum integration as in Eq.(\ref{eq:energydensity}). What this means is that the iterative solution we wrote down is, in reality, an integral equation that we will solve perturbatively in what follows. As mentioned previously, the effective expansion parameter is $\sim \nabla^{2}\phi/H^2$. From the Poisson Eq.(\ref{eq:poisson}) and the Friedmann equation
\begin{equation}
\label{eq:fried}
	H^2 = \frac{8\pi G}{3} \rho_{\textrm{cri}} = \frac{8\pi G}{3} \frac{\bar{\rho}}{\Omega_{\textrm{m}}(a)} \,,
\end{equation}
with $\Omega_{\textrm{m}}(a)= \bar{\rho}(a)/\rho_{\textrm{crit}}(a)$ the fractional contribution of matter to the energy budget, the effective expansion parameter is of order $\nabla^{2}\phi/H^2 \sim \delta = (\rho-\bar{\rho})/\bar{\rho}$ the matter density contrast. Our framework is then an old-fashioned cosmological perturbation theory scheme, in the sense that we can expect it to share the same limitations as traditional methods associated to the fact that the density contrast becomes large at the onset of nonlinearities, signaling the breakdown of the perturbative expansion \cite{Carlson:2009it, Pajer:2017ulp}. In Sec.\ref{sec:eftoflss} we will have more to say about the implications of this observation.
 
\subsection{The perturbative expansion}
\label{sec:pt}

\begin{figure}
\centering
\begin{tikzpicture} 
	\draw	(0,0) -- (0.5,0) (0.75,0) circle(0.25) (1,0) -- (1.5,0);
	\draw [snake] (0.75,0.25) -- (0.75,1);
\end{tikzpicture}
\caption{The only diagram contributing to leading order in the perturbative expansion.}
\label{fig:diagram3}
\end{figure}
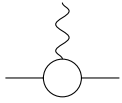

When coupled to the Poisson Eq.(\ref{eq:poisson}), which we here rewrite (in Fourier space) in terms of the Friedmann Eq.(\ref{eq:fried}) evaluated at the present time \footnote{Quantities evaluated at the present time carry a subscript 0. For example, $H_{0}$ is the present day value of the Hubble expansion rate.}
\begin{equation}
\label{eq:poisson2}
	k^2 \phi = -\frac{3}{2} \Omega_{\textrm{m},0}H_{0}^{2} \frac{\delta}{a} \,,
\end{equation}
the iterative (formal) solution we studied before becomes an integral equation that we will here solve in a perturbative expansion. Since the matter density contrast acts as the effective expansion parameter, we look for a solution in the form:
\begin{equation}
\label{eq:pertsol}
	\delta(\eta,\vec{k}) =  \delta^{(1)}(\eta,\vec{k}) + \delta^{(2)}(\eta,\vec{k}) + \cdots \,,
\end{equation}
which in view of Eq.(\ref{eq:poisson2}) translates to a similar expansion for the gravitational potential,
\begin{equation}
	\label{eq:pertsol2}
	\phi(\eta,\vec{k}) =  \phi^{(1)}(\eta,\vec{k}) + \phi^{(2)}(\eta,\vec{k}) + \cdots \,.
\end{equation}
Let us now investigate how the perturbative expansion works explicitly. To leading order the only contribution comes from a single insertion of $\phi^{(1)}$ into the expression for the first order iterative solution in Eq.(\ref{eq:first}). We represent this by the diagram in Fig.\ref{fig:diagram3}, where the number $N$ of wiggly lines connecting to a given circle (in this case $N=1$) determines the order $\phi^{(N)}$ of that insertion \footnote{In general there needs to be at least one wiggly line connecting to any given circle, and the order in perturbation theory can be read from the total number of wiggly lines in a diagram.}.
This diagram evaluates to,
\begin{equation}
\label{eq:11}
	\delta f^{(1)}(\eta, \vec{k}, \vec{q}) = i\vec{k} \cdot \frac{\partial \bar{f}^{(0)}}{\partial \vec{q}} \int_{0}^{\eta} d\eta' a^2(\eta') e^{-i\vec{k}\cdot \vec{q}(\eta-\eta')} \phi^{(1)}(\eta',\vec{k}) \,.
\end{equation}
Also recall that $\bar{f}^{(\textrm{1st})}(\eta,q)=0$ from Eq.(\ref{eq:first}), so the same diagram with a dashed horizontal line vanishes. Now we use the Poisson Eq.(\ref{eq:poisson2}) to write this as
\begin{equation}
	\label{eq:11-2}
	\delta f^{(1)}(\eta, \vec{k}, \vec{q}) = - \frac{i\vec{k}}{k^2} \cdot \frac{\partial \bar{f}^{(0)}}{\partial \vec{q}} \frac{3}{2} \Omega_{\textrm{m},0} H_{0}^{2} \int_{0}^{\eta} d\eta' a(\eta') e^{-i\vec{k}\cdot \vec{q}(\eta-\eta')} \delta^{(1)}(\eta',\vec{k}) \,.
\end{equation}
Next integrate Eq.~(\ref{eq:11-2}) with respect to momentum, using Eq.~(\ref{eq:energydensity}) and the fact that $\bar{f}^{(0)}(q) \propto \delta^{(3)}(\vec{q})$ \footnote{This allows for a straightforward integration over momentum after an integration by parts.}, to arrive at an integral equation for $\delta^{(1)}(\eta,\vec{k})$. One obtains,
\begin{equation}
\label{eq:ie-lo}
	\delta^{(1)}(\eta,\vec{k}) = \frac{3}{2} \Omega_{\textrm{m},0} H_{0}^{2} \int_{0}^{\eta} d\eta' a(\eta') \delta^{(1)}(\eta',\vec{k})(\eta-\eta') \,.
\end{equation}
As expected the linear theory evolution does not couple different wavenumbers, and in fact a separable solution of the form
\begin{equation}
\label{eq:separable}
	\delta^{(1)}(\eta,\vec{k}) = D_{\textrm{L}}(\eta) \delta_{\textrm{L}}(\vec{k}) \,,
\end{equation}
can be found, where $D_{\textrm{L}}(\eta)$ is the linear growth factor (normalized to unity when evaluated today) and $\delta_{\textrm{L}}(\vec{k})$ is the present day linear density field. The latter quantity represents the initial conditions for the matter density contrast, when rescaled to the present time under the assumption of linear evolution. It is a Gaussian stochastic random field whose power spectrum can be extracted from linear Boltzmann solvers (we use the Cosmic Linear Anisotropy Solving System, CLASS \cite{Blas:2011rf}). This provides a precise operational definition for the ensemble averages in our formalism, as they can always be decomposed in terms of the two-point function of the linear density field,
\begin{equation}
\label{eq:2pf}
	\langle \delta_{\textrm{L}}(\vec{k}) \delta_{\textrm{L}}(\vec{k'}) \rangle = (2\pi)^{3} \delta^{(3)}(\vec{k}+\vec{k'}) P_{\textrm{L}}(k) \,,
\end{equation}
with $P_{\textrm{L}}(k)$ the linear theory power spectrum at redshift $z=0$. As we will see, a generic term in the perturbative expansion for the density contrast, in Eq.~(\ref{eq:pertsol}),  scales as $\delta^{(n)} \sim (\delta^{(1)})^{n} = (D_{\textrm{L}} \delta_{\textrm{L}})^{n}$ and so organizes itself in powers of the initial condition for the density contrast.

The evolution equation satisfied by the linear growth factor, Eq.(\ref{eq:ie-lo}), can be recast as a second order ODE by taking two derivatives of this equation with respect to superconformal time
\begin{equation}
\label{ie-lo-ode}
\frac{d^2 D_{\textrm{L}}}{d\eta^2} - \frac{3}{2} \Omega_{\textrm{m},0} H_{0}^{2} a(\eta)D_{\textrm{L}}(\eta) = 0 \,,
\end{equation}
and this looks more familiar when written in terms of the scale factor,
\begin{equation}
\label{eq:ie-lo-ode-time}
	\frac{d^2 D_{\textrm{L}}}{da^2} +\frac{1}{a} \left(3+\frac{d\log H}{d\log a}\right) \frac{dD_{\textrm{L}}}{da} - \frac{3}{2} \Omega_{\textrm{m},0} H_{0}^{2} \frac{D_{\textrm{L}}(a)}{a^5H(a)^2} = 0 \,.
\end{equation}
In Appendix \ref{sec:app1} we derive the well-known analytic solution to this equation, as a special case of the more general scenario involving the presence of a source term on the right-hand side of Eq.~(\ref{eq:ie-lo-ode-time}). There are two independent solutions to this second order ODE 
\begin{equation}
	\label{eq:generalsol_main}
	\begin{split}
		& D_{\textrm{L}}^{+}(a) = H(a) \int_{0}^{a} \frac{da'}{(a')^{3} H^{3}(a')}  \\ &  D_{\textrm{L}}^{-}(a) = H(a) \,.
	\end{split}
\end{equation}
The mode $D_{\textrm{L}}^{-}(a)$ decays with the expansion of the universe, and it quickly becomes negligible in comparison to the growing mode $D_{\textrm{L}}^{+}(a)$. We then drop the decaying mode, and arrive at
\begin{equation}
\label{eq:analytic}
	D_{\textrm{L}}(a) = \frac{H(a)}{H_{0}} \left[\int_{0}^{1} \frac{da'}{(a')^3 H^{3}(a')}\right]^{-1} \int_{0}^{a} \frac{da'}{(a')^3 H^{3}(a')} \,,
\end{equation}  
where we fix the normalization by imposing $D_{\textrm{L}}(a=1)=1$. We also introduce the linear growth rate, $f(a) = d\log D_{\textrm{L}}/d\log a$. Accurate fitting functions for the numerical evaluation of these quantities can be found in \cite{Hamilton:2000tk}.

\begin{figure}
\centering
\begin{tikzpicture} 
	\draw (0.75,0) circle(0.25) (1.75,0) circle(0.25);
	\draw [snake] (0.75,0.25) -- (0.75,1) (1.75,0.25) -- (1.75,1);
	\draw [dashed] (0,0) -- (0.5,0) (1,0) -- (1.5,0)  (2,0) -- (2.5,0);
\end{tikzpicture}
\caption{The only diagram contributing to the background distribution function at second order in perturbation theory.}
\label{fig:diagram5}
\end{figure}
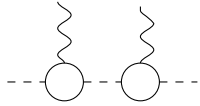

To second order in perturbation theory two diagrams contribute to the distribution function fluctuation as shown in Fig.~\ref{fig:diagram4}. The diagram on the left represents a single insertion of $\phi^{(2)}$ into the first order iterative solution Eq.~(\ref{eq:first}), and the diagram on the right represents two insertions of $\phi^{(1)}$ into the second order iterative solution Eq.~(\ref{eq:second}). For the background distribution function at second order in perturbation theory only a single diagram, as depicted in Fig.\ref{fig:diagram5}, gives a nonzero contribution. 
\begin{figure}
\centering
\begin{tikzpicture} 
 	\draw	(0,0) -- (0.5,0) (0.75,0) circle(0.25) (1,0) -- (1.5,0);
 	\draw (3,0) -- (3.5,0)  (3.75,0) circle(0.25) (4,0) -- (4.5,0) (4.75,0) circle(0.25) (5,0) -- (5.5,0);
 	\draw [snake] (0.75,0.25) -- (0.75,1) (0.75,-0.25) -- (0.75,-1);
 	\draw [snake] (3.75,0.25) -- (3.75,1) (4.75,0.25) -- (4.75,1);
\end{tikzpicture}
\caption{Two diagrams contribute to the distribution function fluctuation at second order in perturbation theory. The total number of wiggly lines reveals the order in perturbation theory. }
\label{fig:diagram4}
\end{figure}
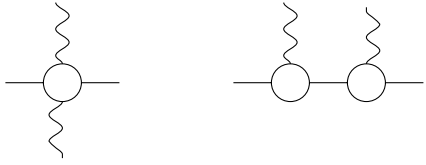

In Appendix \ref{sec:app2} we include detailed calculations of diagrams to second order in perturbation theory, while in the main text we focus on summarizing the main results. The diagram in Fig.\ref{fig:diagram5} evaluates to
\begin{equation}
\label{eq:backrenor}
\begin{split}
	\bar{f}^{(2)}(\eta,q) = - &  \left(\frac{3}{2}\Omega_{\textrm{m},0}H_0^2\right)^2 \int_{0}^{\eta} d\eta' a(\eta') D_{\textrm{L}}(\eta') \int_{0}^{\eta'} d\eta'' a(\eta'') D_{\textrm{L}}(\eta'') \, \times \\ & \times \int \frac{d^3\vec{k}'}{(2\pi)^3} \frac{P_{\textrm{L}}(k')}{(k')^4} i\vec{k}' \cdot \frac{\partial}{\partial \vec{q}} \left[i\vec{k}' \cdot \frac{\partial \bar{f}^{(0)}}{\partial \vec{q}} e^{i\vec{k}' \cdot \vec{q} (\eta'-\eta'')} \right] \,.
\end{split}
\end{equation}
Note that this is a total derivative with respect to momentum. We then see from Eq.(\ref{eq:energydensity}) that this does not lead to a renormalization of the background mass density, which is the statement of particle number conservation. It is also straightforward to check, from Eqs.(\ref{eq:momentum}) and (\ref{eq:backrenor}), that the fluid momentum (and hence the fluid velocity) does not pick up a background renormalization either since $\bar{f}^{(2)}(\eta,q)$ is only a function of the absolute value of comoving momentum, due to statistical isotropy. As a consequence, any perturbative framework based solely on the density and velocity fields necessarily misses this renormalization of the background distribution function. We expect this to remain true at higher orders in the perturbative expansion since particle number conservation and statistical isotropy should hold to all orders. 

That being said, it is straightforward to argue that the background distribution function needs to be renormalized even within SPT. Taking the ensemble average of the stress tensor using Eq.~(\ref{eq:stress}) and the first line of Eq.~(\ref{eq:stresstensor}) yields
\begin{equation}
\label{eq:stressavg}
	\tau(\eta) \equiv \langle \tau^{i}_{i} \rangle\big|_{\eta} = a^{-5}(\eta) \int \frac{d^3 \vec{q}}{(2\pi)^3} q^2 \bar{f}(\eta,q) - \langle \rho v^2 \rangle\big|_{\eta} \,,
\end{equation}
where only the trace can be nonvanishing (upon averaging) due to statistical isotropy. Since the stress tensor vanishes in SPT by construction, the background distribution functions needs to pick up backreactions beyond $\bar{f}^{(0)}(q) \propto \delta^{(3)}(\vec{q})$ in order to cancel the contribution from $\langle \rho v^2 \rangle \neq 0$. Indeed, we show in Appendix \ref{sec:app2} that Eq.(\ref{eq:backrenor}) implies (where we suppress the time dependence for simplicity of notation when it is convenient to do so),
\begin{equation}
\label{eq:stresscancel}
	a^{-5} \int \frac{d^3 \vec{q}}{(2\pi)^3} q^2 \bar{f}^{(2)}(q) = \bar{\rho} \, a^2 H^2 f^2 D_{\textrm{L}}^2 \int_{0}^{\infty} \frac{dk'}{2\pi^2} P_{\textrm{L}}(k') \,.
\end{equation}
This corresponds to the bulk flow (or to be more precise, the linear theory mean square velocity), and exactly cancels the leading contribution to $\langle \rho v^2 \rangle$ in a perturbative expansion, to produce a vanishing averaged stress tensor \footnote{Within SPT our expectation is that higher loop corrections to the background distribution function (of which $\bar{f}^{(4)}(\eta,q)$ is the next to leading order term as we will see shortly) will exactly cancel the higher order contributions to $\langle \rho v^2 \rangle$ such that the consistency relation $\tau(\eta)=0$ is satisfied.}. This is the first indication that our framework based on a perturbative solution to the Vlasov-Poisson system of equations in phase space is reproducing nothing other than SPT.   

Moving on to the distribution function fluctuations, a calculation of the two diagrams in Fig.\ref{fig:diagram4} (carried out explicitly in Appendix \ref{sec:app2}) leads to a second order density contrast which admits a decomposition into a sum of separable terms as follows
\begin{equation}
\label{eq:separable-2nd}
	\delta^{(2)}(a,\vec{k}) = c^{(2)}_{1}(a) h^{(2)}_{1}(\vec{k})+ c^{(2)}_{2}(a) h^{(2)}_{2}(\vec{k}) \,,
\end{equation}
where the scale dependent functions $h^{(2)}_{i}(\vec{k})$ are given by
\begin{equation}
\label{eq:scale}
\begin{split}
	& h^{(2)}_{1}(\vec{k}) = \int_{\{\vec{k}_1, \vec{k}_2\}}^{\vec{k}} \alpha^{(s)}(\vec{k}_1,\vec{k}_2) \normOrd{ \delta_{\textrm{L}}(\vec{k}_1) \delta_{\textrm{L}}(\vec{k}_2)} \\ 	& h^{(2)}_{2}(\vec{k}) = \int_{\{\vec{k}_1, \vec{k}_2\}}^{\vec{k}} \beta(\vec{k}_1,\vec{k}_2) \normOrd{\delta_{\textrm{L}}(\vec{k}_1) \delta_{\textrm{L}}(\vec{k}_2)} \,,
\end{split}
\end{equation}
with $\alpha(\vec{k}_1,\vec{k}_2) = (\vec{k}_1 \cdot \vec{k}_{12})/k_{1}^2$ and $\beta(\vec{k}_1,\vec{k}_2) = k_{12}^2(\vec{k}_1 \cdot \vec{k}_2)/2k_1^2 k_2^2$. Here $\vec{k}_{12}=\vec{k}_1 + \vec{k}_2$ and $\alpha^{(s)}(\vec{k}_1,\vec{k}_2) = [\alpha(\vec{k}_1,\vec{k}_2) + \alpha(\vec{k}_2,\vec{k}_1)]/2$ stands for the symmetric combination. We derive explicit analytic expressions for the time-dependent coefficients, $c^{(2)}_{i}(a)$, in Appendix \ref{sec:app1}. 

Let us now move on to the third order in the perturbative expansion. The background distribution function picks up no third order contributions, or to any odd order in perturbation theory more generally, due to the assumption of a Gaussian linear random field. In pictorial language, all diagrams with a horizontal dashed line and an odd number of wiggly lines vanish. There are, however, four distinct diagrams contributing to the distribution function fluctuation at third order in perturbation theory, as illustrated in Fig.\ref{fig:diagram6}. After evaluating these diagrams, we find that the third order density contrast can also be decomposed as a sum of separable terms \footnote{One can organize such a decomposition into a proper basis of scale dependent functions \cite{Hartmeier:2023brx}, but we are not going to worry about this here.}
\begin{figure}
	\centering
	\begin{tikzpicture} 
		\draw (0,0) -- (0.5,0) (0.75,0) circle(0.25) (1,0) -- (1.5,0) (2,0) -- (2.5,0) (2.75,0) circle (0.25) (3,0) -- (3.5,0) (3.75,0) circle (0.25) (4,0) -- (4.5,0);
		\draw [snake] (0.75, -0.25) -- (0.75, -1) (0.5335, 0.125) -- (-0.1160, 0.5);
		\draw [snake] (0.9665, 0.125) -- (1.6160, 0.5);
		\draw [snake]  (2.75, 0.25) -- (2.75, 1) (2.75,-0.25) -- (2.75,-1);
		\draw [snake] (3.75, 0.25) -- (3.75,1);
		\draw (5,0) -- (5.5,0) (5.75,0) circle(0.25) (6,0) -- (6.5,0) (6.75,0) circle(0.25) (7,0) -- (7.5,0);
		\draw [snake] (5.75,0.25) -- (5.75, 1);
		\draw [snake]  (6.75,0.25) -- (6.75, 1) (6.75,-0.25) -- (6.75,-1);
		\draw (8,0) -- (8.5,0) (8.75,0) circle(0.25) (9,0) -- (9.5,0) (9.75,0) circle(0.25) (10,0) -- (10.5,0) (10.75, 0) circle(0.25) (11,0) -- (11.5,0);
		\draw [snake] (8.75,0.25) -- (8.75,1);
		\draw [snake]  (9.75, 0.25) -- (9.75, 1);
		\draw [snake] (10.75, 0.25) -- (10.75, 1);
	\end{tikzpicture}
	\caption{Four diagrams contribute to the distribution function fluctuation at third order in perturbation theory.}
	\label{fig:diagram6}
\end{figure}

\begin{equation}
	\label{eq:separable-3rd}
	\delta^{(3)}(a, \vec{k}) = \sum_{i=1}^{6} c^{(3)}_{i}(a) h^{(3)}_{i}(\vec{k}) \,.
\end{equation}
The scale dependent functions are,
\begin{equation}
	\label{eq:scale-3rd}
	\begin{split}
		& h^{(3)}_{1}(\vec{k}) = \int_{\{\vec{k}_1, \vec{k}_2, \vec{k}_3\} }^{\vec{k}} \alpha(\vec{k}_1,\vec{k}_{23})\alpha^{(s)}(\vec{k}_2,\vec{k}_3) \, \delta_{\textrm{L}}(\vec{k}_1) \normOrd{ \delta_{\textrm{L}}(\vec{k}_2) \delta_{\textrm{L}}(\vec{k}_3)} \\ 	& h^{(3)}_{2}(\vec{k}) = \int_{\{\vec{k}_1, \vec{k}_2, \vec{k}_3\} }^{\vec{k}} \alpha(\vec{k}_1,\vec{k}_{23})\beta(\vec{k}_2,\vec{k}_3) \, \delta_{\textrm{L}}(\vec{k}_1) \normOrd{ \delta_{\textrm{L}}(\vec{k}_2) \delta_{\textrm{L}}(\vec{k}_3)} \\ &  h^{(3)}_{3}(\vec{k}) = \int_{\{\vec{k}_1, \vec{k}_2, \vec{k}_3\} }^{\vec{k}} \alpha(\vec{k}_{23},\vec{k}_1)\alpha^{(s)}(\vec{k}_2,\vec{k}_3) \, \delta_{\textrm{L}}(\vec{k}_1) \normOrd{ \delta_{\textrm{L}}(\vec{k}_2) \delta_{\textrm{L}}(\vec{k}_3)} \\ & h^{(3)}_{4}(\vec{k}) = \int_{\{\vec{k}_1, \vec{k}_2, \vec{k}_3\} }^{\vec{k}} \alpha(\vec{k}_{23},\vec{k}_1)\beta(\vec{k}_2,\vec{k}_3) \, \delta_{\textrm{L}}(\vec{k}_1) \normOrd{ \delta_{\textrm{L}}(\vec{k}_2) \delta_{\textrm{L}}(\vec{k}_3)} \\ & h^{(3)}_{5}(\vec{k}) = \int_{\{\vec{k}_1, \vec{k}_2, \vec{k}_3\} }^{\vec{k}} \beta(\vec{k}_1,\vec{k}_{23})\alpha^{(s)}(\vec{k}_2,\vec{k}_3) \, \delta_{\textrm{L}}(\vec{k}_1) \normOrd{ \delta_{\textrm{L}}(\vec{k}_2) \delta_{\textrm{L}}(\vec{k}_3)} \\ & h^{(3)}_{6}(\vec{k}) = \int_{\{\vec{k}_1, \vec{k}_2, \vec{k}_3\} }^{\vec{k}} \beta(\vec{k}_1,\vec{k}_{23})\beta(\vec{k}_2,\vec{k}_3) \, \delta_{\textrm{L}}(\vec{k}_1) \normOrd{ \delta_{\textrm{L}}(\vec{k}_2) \delta_{\textrm{L}}(\vec{k}_3)} \,,
	\end{split}
\end{equation}
and analytic formulas for the time-dependent coefficients, $c_{i}^{(3)}(a)$, are provided in Appendix \ref{sec:app2}. 

Our results in this section are in exact agreement with the SPT predictions for the density contrast, obtained following the traditional approach outlined in Sec.\ref{sec:spt} (see, e.g., \cite{Hartmeier:2023brx, Takahashi:2008yk, Bernardeau:1993qu, Fasiello:2022lff, Fasiello:2016qpn, Choustikov:2023uyk, Lewandowski:2016yce, Fujita:2020xtd, Donath:2020abv, Schmidt:2020ovm, Rampf:2022tpg, Garny:2022fsh})\footnote{Note that normal ordering symbols appear in Eqs.~(\ref{eq:scale}) and (\ref{eq:scale-3rd}) for the perturbation theory kernels. However, in SPT one obtains the same kernels without any normal ordering symbols. This is inconsequential as the forbidden contractions would lead to a vanishing contribution. The normal ordering symbols then simply organize for us which of the contractions lead to a nonvanishing contribution and hence need to be evaluated in the first place.}. However, let us stress that at no point in our framework did we have to assume a vanishing stress tensor or vorticity. Instead, an ideal pressureless fluid follows as a consequence of the perturbative solution to the Vlasov-Poisson system of equations in phase space. This is consistent with the intuition that dissipative effects are due to intrisically nonperturbative effects in nonlinear gravitational evolution. We attribute the following physical interpretation to this result: Our perturbative expansion is based on the iterative approach developed in Sec.\ref{sec:iter}, which reconstructs particle trajectories in an expansion in powers of the gradient of the gravitational field, as first showed in \cite{Senatore:2017hyk}. Now, the shell-crossing singularity occurs when particles start to accumulate in small regions of space, which require particle trajectories to turn around. This happens when the gradient of the gravitational field becomes sufficiently large, likely beyond the convergence radius of the iterative solution. This is why the perturbative expansion does not account for the shell-crossing singularity.  

This result is consistent with the findings of \cite{Valageas:2001df}, where the author offers the following explanation as to why the perturbative expansion based on the Vlasov-Poisson system reproduces SPT: At early times, before the onset of shell-crossing, the fluid description of the system holds exactly and hence must match what is obtained from the Boltzmann equation. Since the structure of the perturbative expansion at all times is set in terms of powers of the initial conditions via $\delta_{\textrm{L}}(k)$, there is no room to accommodate for shell-crossing at late times once it first occurs. Beyond this simple argument, there are no definitive proofs that the perturbative expansion based on the Vlasov-Poisson system exactly reproduces SPT to all orders in perturbation theory \footnote{In particular, we were not able to find a symmetry argument which manifestly forbids the generation of a nonzero average stress tensor, i.e., which explains the exact cancellation of diagrams identified in Eqs.~(\ref{eq:stressavg}) and (\ref{eq:stresscancel}).}.
	
The formalism we develop in this paper is an improvement over the one introduced in \cite{Valageas:2001df}, mainly for the added clarity on how the nonlinearities in gravitational evolution renormalize the background distribution function. This feature enables us to make a connection between the cosmological perturbation theory in phase space and the EFTofLSS, which will be the subject of our attention in the next section.

We intend Sec.\ref{sec:ptbe} to be a useful pedagogical reference for numerical calculations in SPT with the full $\Lambda$CDM time-dependent kernels, up to third order in the perturbative expansion. For this reason, we supplement the main text with explicit analytic formulas for the time-dependent coefficients in Appendix \ref{sec:app1}, one-loop power spectrum in Appendix \ref{sec:app4} and tree-level bispectrum in Appendix \ref{sec:app5}.

\section{Connection to EFT methods}
\label{sec:eftoflss}

In Sec.~\ref{sec:ptbe} we showed that the Boltzmann equation can be recast as a coupled set of equations, one for the background distribution function, Eq.~(\ref{eq:vlasovavg}), and another for its fluctuations, Eq.~(\ref{eq:vlasovfluc}). We then solved these two coupled set of equations perturbatively in full phase space, while also imposing the Poisson Eq.(\ref{eq:poisson2}). We saw that the outcome of this procedure is identical to SPT, in the sense that both frameworks lead to the same density contrast at second and third order in perturbation theory. 

The key difference is that our perturbative framework based on the Vlasov-Poisson system of equations in phase space is based solely on the most fundamental object, the distribution function. In fact, one important outcome of our study was the observation that the background distribution function picks up backreactions from gravitational nonlinearities, a feature that is not transparent in the traditional approach based on the density and velocity fields alone since these quantities do not get renormalized. We expect such backreactions into the background distribution function to arise not only from perturbative scales, but also from nonperturbative ones. For example, within halos the average distribution function was found to have a Gaussian core with exponential wings due to virial velocities \cite{Bryan_1998, Sheth:2000ii}. This observation suggests a natural path towards improving on SPT: To insert the \textit{apriori} unknown fully nonlinear background distribution function into Eq.(\ref{eq:vlasovfluc}) for its fluctuations, which should then be solved perturbatively as before.

We begin this section by exploring that idea, which will naturally point at a connection to EFT methods. We will see that the sound speed counterterm naturally emerges from the theory even without EFT ingredients, but we will ultimately argue that EFT methods are strictly necessary to account for a nonzero velocity dispersion in a fully self-consistent framework.

\subsection{Emergence of the counterterm}
\label{sec:resbdf}

Consider a split of the fully nonlinear background distribution function, 
\begin{equation}
\label{eq:distres}
	\bar{f}(\eta, q) = \bar{f}_{\textrm{P}}(\eta, q) + \bar{f}_{\textrm{ctr}}(\eta,q) \,,
\end{equation}
into a perturbative piece, $\bar{f}_{\textrm{P}}(\eta,q)$, which we calculated to third order in perturbation theory in Sec.\ref{sec:ptbe} [see Eq.(\ref{eq:backrenor}), and recall that odd order terms vanish]
\begin{equation}
\label{eq:distpert}
	\bar{f}_{\textrm{P}}(\eta, q) = \bar{f}_{\textrm{P}}^{(0)}(q) + \bar{f}_{\textrm{P}}^{(2)}(\eta, q) + \cdots \,,
\end{equation}
and another one which we refer to as a counterterm piece, $\bar{f}_{\textrm{ctr}}(\eta,q)$, with the benefit of hindsight. It accounts for the backreactions from short distance scales that are not under perturbative control,
\begin{equation}
\label{eq:distresum2}
	\bar{f}_{\textrm{ctr}}(\eta,q) = \bar{f}_{\textrm{ctr}}^{(2)}(\eta, q) + \cdots \,.
\end{equation}
We take the leading contribution to $\bar{f}_{\textrm{ctr}}(\eta,q)$ as a second order quantity in perturbation theory since it is sourced by quadratic nonlinearities \footnote{This is not strictly necessary as one can remain agnostic about the size of $\bar{f}_{\textrm{ctr}}(\eta,q)$, which can be thought of as resuming the dissipative effects associated to a nonzero average velocity dispersion to all orders. In that case, deviations from an ideal pressureless fluid appear already at the linear theory level. This choice is made, for instance, in \cite{Garny:2022tlk, Garny:2022kbk, Garny:2025ovs}. On the other hand, the standard EFTofLSS assumes its counterterms to have a leading second order contribution as in our framework, \label{footnote1} so that linear theory is not modified.}.

We now proceed to investigate the contribution from the new counterterm piece of the background distribution function to the perturbative expansion of the fluctuation $\delta f(\eta, \vec{k},\vec{q})$. As we have shown in the previous section, it is only at third order in perturbation theory that the background distribution function backreacts into the fluctuations as given by the first line in Eq.~(\ref{eq:third2}), and due to the presence of $\bar{f}_{\textrm{ctr}}^{(2)}(\eta, q)$ there will be a new term contributing to the integral equation that needs to be solved at third order. Its diagrammatic representation can be found in Fig.\ref{fig:diagram7}, and it evaluates to
\begin{figure}[h]
	\centering
	\begin{tikzpicture} 
		\draw	(0,0) -- (0.5,0) (0.75,0) circle(0.25) (1,0) -- (1.5,0);
		\draw [snake] (0.75,0.25) -- (0.75,1);
		\draw  (1.323, -0.176) -- (1.677, 0.176) (1.323, 0.176) -- (1.677, -0.176);
	\end{tikzpicture}
	\caption{Diagram for the counterterm background distribution function contribution to fluctuations at third order in perturbation theory.}
	\label{fig:diagram7}
\end{figure}
\begin{equation}
\label{eq:resum1}
	\delta f^{(3)}(\eta, \vec{k},\vec{q}) = \int_{0}^{\eta} d\eta' a^2(\eta') e^{-i\vec{k} \cdot \vec{q} (\eta-\eta')} \phi^{(1)}(\eta',\vec{k}) i\vec{k} \cdot \frac{\partial \bar{f}_{\textrm{ctr}}^{(2)}}{\partial \vec{q}} \Bigg|_{\eta',q} + \cdots \,,
\end{equation}
where the ellipsis represent all perturbative contributions which were accounted for in the previous subsection. Using the Poisson Eq.~(\ref{eq:poisson2}), Eq.~(\ref{eq:separable}) and integrating over momentum to obtain the density perturbation as in Eq.~(\ref{eq:energydensity}) yields
\begin{equation}
\label{eq:resum2}
\begin{split}
	\delta^{(3)}(\eta,\vec{k},\vec{q}) = & \frac{1}{a^3\bar{\rho}} \delta_{\textrm{L}}(\vec{k}) \ \frac{3}{2} \Omega_{\textrm{m},0} H_{0}^2 \int_{0}^{\eta} d\eta' a(\eta') D_{\textrm{L}}(\eta')(\eta-\eta') \times \\ & \times \int \frac{d^3 \vec{q}}{(2\pi)^3} e^{-i\vec{k} \cdot \vec{q} (\eta-\eta')} \bar{f}_{\textrm{ctr}}^{(2)}(\eta',q) + \cdots \,.
\end{split}
\end{equation}
Let us take a closer look into the integral over momentum in the second line of Eq.~(\ref{eq:resum2}). First recall that the comoving momentum is defined by $\vec{q}=a^2(d\vec{x}/dt)$ and hence  $q \int dt/a^2 \sim q\eta$ is the comoving distance traveled by cold dark matter particles, which is known to be of order $\sim 1/k_{\textrm{NL}}$ with $k_{\textrm{NL}}$ the scale of nonlinearities. This is the physical scale above which one expects SPT to break down, and as a consequence perturbative methods efficiently model the dynamics on scales $k \ll k_{\textrm{NL}}$. Now note that the argument of the exponent in the second line of Eq.~(\ref{eq:resum2}) scales like $\sim k/k_{\textrm{NL}} \ll 1$, which justifies a Taylor series expansion
\begin{equation}
\label{eq:taylor}
	e^{-i\vec{k} \cdot \vec{q} (\eta-\eta')} = 1 -i\vec{k} \cdot \vec{q} (\eta-\eta') - \frac{1}{2} (\vec{k} \cdot \vec{q})^2 (\eta-\eta')^2 + \cdots \,.
\end{equation} 
Here the zeroth order term corresponds to a renormalization of the background mass density, which we know does not occur in perturbation theory from Sec.~\ref{sec:pt} (we will show this explicitly for the counterterm piece as well in Sec.~\ref{sec:eft}, within the context of EFT methods). Also, the first order term is proportional to $\sim (\hat{k} \cdot \hat{q})$ which vanishes after integrating over the solid angle. We are then left with
\begin{equation}
\label{eq:taylor2}
	\int \frac{d^3 \vec{q}}{(2\pi)^3} e^{-i\vec{k} \cdot \vec{q} (\eta-\eta')} \bar{f}_{\textrm{ctr}}^{(2)}(\eta',q) \approx  -\frac{1}{2} (\eta-\eta')^2 \int \frac{d^3 \vec{q}}{(2\pi)^3} (\vec{k} \cdot \vec{q})^2 \bar{f}_{\textrm{ctr}}^{(2)}(\eta',q)  \,, 
\end{equation}
After integrating over the solid angle, the substitution of Eq.~(\ref{eq:taylor2}) into Eq.~(\ref{eq:resum2}) gives 
\begin{equation}
\label{eq:resum3}
	\delta^{(3)}(\eta,\vec{k},\vec{q}) = -\frac{1}{6} k^2  \delta_{\textrm{L}}(\vec{k}) \times \frac{3}{2} \Omega_{\textrm{m},0} H_{0}^2 \int_{0}^{\eta} d\eta' a^{3}(\eta') D_{\textrm{L}}(\eta') (\eta-\eta')^3 \sigma^2_{\textrm{dis}}(\eta') + \cdots \,, \small
\end{equation}
where we introduced the average velocity dispersion squared
\begin{equation}
\label{eq:resvelocity}
\bar{\rho}(\eta) \sigma^2_{\textrm{dis}}(\eta) = a^{-5}(\eta) \int \frac{d^3 \vec{q}}{(2\pi)^3} q^2 \bar{f}_{\textrm{ctr}}^{(2)}(\eta,q) \,,
\end{equation}
and used the relation $a^3(\eta)\bar{\rho}(\eta) = a^3(\eta') \bar{\rho}(\eta')$ to bring this quantity inside the time integral. 

Before moving forward, let us stop for a moment to see explicitly that Eq.~(\ref{eq:resvelocity}) indeed corresponds to the average velocity dispersion, a quantity directly related to the averaged stress tensor $\tau(\eta)$, which reads from Eq.~(\ref{eq:stressavg})
\begin{equation}
	\label{eq:stressavgagain}
	\bar{\rho}(\eta) \sigma^2_{\textrm{dis}}(\eta) \equiv \tau(\eta) = a^{-5}(\eta) \int \frac{d^3 \vec{q}}{(2\pi)^3} q^2 \bar{f}(\eta,q) - \langle \rho v^2 \rangle\big|_{\eta} \,.
\end{equation}
However, from Eq.~(\ref{eq:distres}) we have that $\bar{f}(\eta, q) = \bar{f}_{\textrm{P}}(\eta, q) + \bar{f}_{\textrm{ctr}}(\eta,q)$, and the contributions from $\bar{f}_{\textrm{P}}(\eta, q)$ exactly cancel those from $\langle \rho v^2 \rangle\big|_{\eta}$ according to the arguments made around Eq.~(\ref{eq:stresscancel}) in Sec.~\ref{sec:pt} (also see a discussion about this result at the end of that section). We are then left with Eq.~(\ref{eq:resvelocity}).

The diagram in Fig.~\ref{fig:diagram7}, which evaluates to Eq.~(\ref{eq:resum3}), introduces an extra contribution, $c^{(3)}_{\textrm{ctr}}(a) h^{(3)}_{\textrm{ctr}}(\vec{k})$, to the decomposition of the third order density contrast into a sum of separable terms,  Eq.~(\ref{eq:separable-3rd}). The scale dependent part can be read off to be,
\begin{equation}
\label{eq:scale-res}
	h_{\textrm{ctr}}(\vec{k}) = -\frac{1}{2} k^2 \delta_{\textrm{L}}(\vec{k}) \,.
\end{equation}
The time-dependent coefficient, $c^{(3)}_{\textrm{ctr}}(a)$, can be computed using the machinery developed in Appendices \ref{sec:app1} and \ref{sec:app2}. We summarize here the steps involved for completeness. We first need to compute the source function $s^{(3)}_{\textrm{ctr}}(a)$, associated to $c^{(3)}_{\textrm{ctr}}(a)$ via Eq.~(\ref{eq:sourced}). As explained in Appendix \ref{sec:app2}, this can be obtained from Eq.~(\ref{eq:resum3}) by differentiating it twice with respect to superconformal time followed by a division of the result by a factor of $a^6 H^2$. This reads [also factoring out $h^{(3)}_{\textrm{ctr}}(\vec{k})$ from Eq.~(\ref{eq:resum3})]:
\begin{equation}
	\label{eq:source-re}
	s^{(3)}_{\textrm{ctr}} = \frac{2}{a^6H^2} \ \frac{3}{2} \Omega_{\textrm{m},0} H_{0}^2 \int_{0}^{\eta} d\eta' a^{3}(\eta') D_{\textrm{L}}(\eta')(\eta-\eta')	\sigma^2_{\textrm{dis}}(\eta') \,.
\end{equation} 
The time-dependent coefficient $c^{(3)}_{\textrm{ctr}}(a)$ can then be computed from Eq.~(\ref{eq:analyticsolagainapp}) with the source term Eq.~(\ref{eq:source-re}), assuming knowledge of $\bar{f}_{\textrm{ctr}}^{(2)} (q,\eta)$ and hence of $\sigma^2_{\textrm{dis}}(\eta)$. It then follows from Eq.(\ref{eq:scale-res}) that this new contribution to the density contrast at third order, coming from the backreaction of the counterterm background distribution function into its fluctuations, adds a new term to the one-loop power spectrum
\begin{equation}
\label{eq:powerres}
	\Delta P(a,k)  = P_{\textrm{1-loop}}(a,k)- P_{\textrm{1-loop,SPT}}(a,k) = -D_{\textrm{L}}(a)c^{(3)}_{\textrm{ctr}}(a)k^2 P_{\textrm{L}}(k) \,.
\end{equation}

This accounts for a nonzero average velocity dispersion, and it has the exact same form as the effective sound speed counterterm in the EFTofLSS. In fact, our framework qualitatively captures the same physical effects as the EFT: The backreation of short distance fluctuations into the background, and its impact on the dynamics of long-wavelength fluctuations \cite{Baumann:2010tm, Carrasco:2012cv}. 

Since $\bar{f}_{\textrm{ctr}}^{(2)} (q,\eta)$ is not a priori known, the same is true of $\sigma^2_{\textrm{dis}}(\eta)$ and hence of $c^{(3)}_{\textrm{ctr}}(a)$ as well. Instead, one can think of it as a free parameter to be determined by matching to either N-body simulations or observations. This is analogous to the case of chiral perturbation theory in QCD \cite{Gasser:1983yg}, where a nonzero quark condensate emerges due to nonperturbative effects (i.e. confinement)  and the underlying EFT is build from an apriori unknown value for this quantity. Also note that the scale dependence in Eq.~(\ref{eq:powerres}) is just right so that the free coefficient $c^{(3)}_{\textrm{ctr}}(a)$ can absorb UV divergences present in the perturbation theory loop integrals, as is well-known in the EFTofLSS \footnote{Stochastic terms are also required to absorb all divergences. We will have more to say about stochastic terms in the next section.}. One can then proceed to renormalize cosmological perturbation theory in the exact same way one would any other field theory, for cosmologies that suffer from those UV divergences \cite{Pajer:2013jj}.

So long as we think of $c^{(3)}_{\textrm{ctr}}(a)$ as a free parameter, our formalism leads to a model for the one-loop power spectrum witch is identical to the EFTofLSS in its simplest form. This is not, however, the end of the story. The reason for this is the fact that we do happen to have a really good handle on the nonperturbative gravitational dynamics from first principles via N-body simulations (as opposed to the more traditional EFT approach to parametrize unknown physics), so we should be able to extract the counterterm from simulations  and, crucially, have a physical interpretation for what it represents. The analogy to chiral perturbation theory, which has the pion decay constant as a free parameter, makes sense here as well. This quantity can be extracted from first principles using QCD lattice simulations  \cite{Wilson:1974sk, McNeile:2006qy, Fukugita:1992np, Mastropas:2014fsa} and has a clear physical interpretation.

Previous works have pointed out that SPT can be improved by incorporating a nonzero average velocity dispersion in the formalism. It was argued in \cite{Pueblas:2008uv, Erschfeld:2018zqg} that a nonzero average velocity dispersion can regulate the shell-crossing singularity, and our setup is similar to the approaches taken in both \cite{Garny:2022tlk,Garny:2022kbk,Garny:2025ovs} and \cite{Aviles:2015osc}, where a nonzero average velocity dispersion is incorporated in the perturbative expansion, leading to the emergence of EFT-like counterterms. We reproduce these results within a framework that solves the Vlasov-Poisson system of equations directly in phase space, and hence does not require a truncation of the Boltzmann hierarchy from the outset. 

The results obtained in this section seem to provide a physical interpretation for the counterterm in terms of an average velocity dispersion. In what follows we show, however, that this is necessarily incomplete due to short distance fluctuations which are not under perturbative control. To be concrete, we will see that a nonzero average velocity dispersion is directly sourced by a short-scale gravitational binding energy which is not accounted for in our framework, but it should contribute to the counterterm. On top of that, it is not the averaged quantities that source the EFT counterterm, but rather their response to the presence of a long-wavelength mode (see \cite{Nascimento:2024rbv} for a concrete realization of this within the separate universe picture). For these reasons, the EFT approach is strictly necessary to account for a nonzero velocity dispersion within a fully self-consistent framework.

\subsection{EFTofLSS framework}
\label{sec:eft}

We now introduce standard EFT techniques to make the connection between the formalism developed in Sec.~\ref{sec:resbdf} and the EFTofLSS more precise. As a starting point, let us repeat here the collisionless Boltzmann (Vlasov) Eq.~(\ref{eq:vlasov3}),
\begin{equation}
	\label{eq:vlasovagain}
	\frac{\partial f}{\partial \eta} + \vec{q} \cdot \frac{\partial f}{\partial \vec{x}} =a^2(\eta) \frac{\partial \phi}{\partial \vec{x}} \cdot \frac{\partial f}{\partial \vec{q}} \,.
\end{equation} 
The nonlinear term on the right-hand side of Eq.~(\ref{eq:vlasovagain}) is a contact term \footnote{A product of two fields evaluated at the same point in space.}, being sensitive to nonlinearities in gravitational dynamics at arbitrarily small scales. In order to have theoretical control over scales that cannot be treated perturbatively, we adopt standard EFT techniques and split the distribution function into a long-wavelength piece, defined by smoothing over a (distance) scale $1/\Lambda$, 
\begin{equation}
\label{eq:smoothing}
	f_{l}(\eta,\vec{x},\vec{q}) = \int d^{3} \vec{x}' \, W_{\Lambda}(|\vec{x}-\vec{x}'|)f(\eta,\vec{x}',\vec{q}) \,,
\end{equation}
and a short-wavelength piece $f_{s}=f-f_{l}$ \footnote{A similar decomposition, $\phi = \phi_{l}+\phi_{s}$, also follows for the gravitational potential, since it is linearly related to the distribution function by Eqs.~(\ref{eq:energydensity}) and (\ref{eq:poisson}).}. The evolution equation for the long-wavelength distribution function follows from Eqs.~(\ref{eq:vlasovagain}) and (\ref{eq:smoothing})
 \begin{equation}
 \label{eq:vlasovlong}
 	\frac{\partial f_{l}}{\partial \eta} + \vec{q} \cdot \frac{\partial f_{l}}{\partial \vec{x}} = a^{2}(\eta) \int d^{3}\vec{x}' \, W_{\Lambda}(|\vec{x}-\vec{x}'|) \frac{\partial \phi (\eta,\vec{x}',\vec{q})}{\partial \vec{x}'} \cdot \frac{\partial f(\eta,\vec{x}',\vec{q})}{\partial \vec{q}} \,.
\end{equation}
By following standard methods (e.g., see \cite{Baumann:2010tm}) one obtains the result that mixed terms involving a product of long and short modes are suppressed by powers of $(k/\Lambda)^2 \ll 1$, and we arrive at
\begin{equation}
	\label{eq:vlasovlong2}
	\frac{\partial f_{l}}{\partial \eta} + \vec{q} \cdot \frac{\partial f_{l}}{\partial \vec{x}} = a^{2}(\eta) \frac{\partial \phi_{l}}{\partial \vec{x}} \cdot \frac{\partial f_{l}}{\partial \vec{q}} + a^2(\eta) \left[\frac{\partial \phi_{s}}{\partial \vec{x}} \cdot \frac{\partial f_{s}}{\partial \vec{q}} \right]_{\Lambda} + O\left(\frac{k^2}{\Lambda^2}\right) \,,
\end{equation}
where,
\begin{equation}
\label{eq:shortmodes}
	\left[\frac{\partial \phi_{s}}{\partial \vec{x}} \cdot \frac{\partial f_{s}}{\partial \vec{q}} \right]_{\Lambda} = \int d^3 \vec{x}' \, W_{\Lambda}(|\vec{x}-\vec{x}'|) \frac{\partial \phi_{s} (\eta,\vec{x}',\vec{q})}{\partial \vec{x}'} \cdot \frac{\partial f_{s}(\eta,\vec{x}',\vec{q})}{\partial \vec{q}} \,.
\end{equation}
 
To make direct contact with the formalism developed in Sec.~\ref{sec:ptbe} we first split the Boltzmann equation into two coupled set of equations, one for the background distribution function and another for its fluctuation. The first term on the right-hand side of Eq.(\ref{eq:vlasovlong2}) can be decomposed as before
\begin{equation}
\label{eq:decomplong}
	a^{2}(\eta) \frac{\partial \phi_{l}}{\partial \vec{x}} \cdot \frac{\partial f_{l}}{\partial \vec{q}} = 	a^{2}(\eta) \normOrd{\frac{\partial \phi_{l}}{\partial \vec{x}} \cdot \frac{\partial f_{l}}{\partial \vec{q}}} + \ a^{2}(\eta) \left\langle \frac{\partial \phi_{l}}{\partial \vec{x}} \cdot \frac{\partial f_{l}}{\partial \vec{q}} \right\rangle \,,
\end{equation}
where the normal ordering symbol subtracts averages as defined in Eq.~(\ref{eq:normalordering}). A similar expression holds for the second term on the right-hand side of Eq.~(\ref{eq:vlasovlong2})
\begin{equation}
\label{eq:decomshort}
	 a^2(\eta) \left[\frac{\partial \phi_{s}}{\partial \vec{x}} \cdot \frac{\partial f_{s}}{\partial \vec{q}} \right]_{\Lambda} =  a^2(\eta) \normOrd{ \left[\frac{\partial \phi_{s}}{\partial \vec{x}} \cdot \frac{\partial f_{s}}{\partial \vec{q}} \right]_{\Lambda}} + a^2(\eta) \left\langle \left[\frac{\partial \phi_{s}}{\partial \vec{x}} \cdot \frac{\partial f_{s}}{\partial \vec{q}} \right]_{\Lambda} \right\rangle  \,.
\end{equation}
From Eq.~(\ref{eq:shortmodes}), the second term on the right-hand side of Eq.~(\ref{eq:decomshort}) simplifies to
\begin{equation}
\label{eq:average_short}
	\left\langle \left[\frac{\partial \phi_{s}}{\partial \vec{x}} \cdot \frac{\partial f_{s}}{\partial \vec{q}} \right]_{\Lambda} \right\rangle = \left\langle \frac{\partial \phi_{s}}{\partial \vec{x}} \cdot \frac{\partial f_{s}}{\partial \vec{q}} \right\rangle \,.
\end{equation}
We can also rewrite the first term on the right-hand side of Eq.~(\ref{eq:decomshort}) as follows. First define the stochastic term,
\begin{equation}
\label{eq:stochastic}
		\epsilon_{\Lambda} \equiv \left[\frac{\partial \phi_{s}}{\partial \vec{x}} \cdot \frac{\partial f_{s}}{\partial \vec{q}} \right]_{\Lambda} -  \left\langle \left[\frac{\partial \phi_{s}}{\partial \vec{x}} \cdot \frac{\partial f_{s}}{\partial \vec{q}} \right]_{\Lambda} \right\rangle_{\delta f_{l}(\eta,\vec{x},\vec{q})} \,
\end{equation}
where the subscript $\delta f_{l}(\eta,\vec{x},\vec{q})$ on the second term in the previous expression indicates that the average is to be taken on the presence of a long-wavelength fluctuation to the distribution function. The quantity $\epsilon_{\Lambda}$ is entirely analogous to the stochastic term usually introduced in the context of the standard EFTofLSS framework \cite{Baumann:2010tm, Carrasco:2012cv}. Upon expanding the short scale fluctuations in terms of the long-wavelength mode, we obtain
\begin{equation}
\label{eq:short_expansion}
	\left\langle \left[\frac{\partial \phi_{s}}{\partial \vec{x}} \cdot \frac{\partial f_{s}}{\partial \vec{q}} \right]_{\Lambda} \right\rangle_{\delta f_{l}(\eta,\vec{x},\vec{q})} - \left\langle \frac{\partial \phi_{s}}{\partial \vec{x}} \cdot \frac{\partial f_{s}}{\partial \vec{q}} \right\rangle = \frac{\partial}{\partial \delta f_{l}}  	\left\langle \left[\frac{\partial \phi_{s}}{\partial \vec{x}} \cdot \frac{\partial f_{s}}{\partial \vec{q}} \right]_{\Lambda} \right\rangle_{\delta f_{l}} \Bigg|_{\delta f_{l}=0} \delta f_{l}(\eta,\vec{x},\vec{q}) + \cdots \,,
\end{equation} 
where the ellipsis represent terms of higher order in $\delta f_{l}$ and derivatives thereof. Combining Eqs.~(\ref{eq:decomshort}), (\ref{eq:stochastic}) and (\ref{eq:short_expansion}) now yields
\begin{equation}
\label{eq:normalorder_short}
	\normOrd{ \left[\frac{\partial \phi_{s}}{\partial \vec{x}} \cdot \frac{\partial f_{s}}{\partial \vec{q}} \right]_{\Lambda}} = \frac{\partial}{\partial \delta f_{l}}  	\left\langle \left[\frac{\partial \phi_{s}}{\partial \vec{x}} \cdot \frac{\partial f_{s}}{\partial \vec{q}} \right]_{\Lambda} \right\rangle_{\delta f_{l}} \Bigg|_{\delta f_{l}=0} \delta f_{l}(\eta,\vec{x},\vec{q}) + \cdots + \epsilon_{\Lambda}
\end{equation}

Going back to the Vlasov Eq.~(\ref{eq:vlasovlong2}), its ensemble average reads \footnote{Note that we drop higher derivative terms of $O(k^2/\Lambda^2)$ from now on. However, it is important to keep in mind that such terms are necessary to accurately model the nonlinear power spectrum for any finite value of the cutoff $\Lambda$ \cite{Baumann:2010tm, Carrasco:2012cv}.}
\begin{equation}
\label{eq:avgvlasov}
	\frac{\partial \bar{f}_{l}}{\partial \eta} = a^2(\eta) \left\langle \frac{\partial \phi_{l}}{\partial \vec{x}} \cdot \frac{\partial f_{l}}{\partial \vec{q}} \right \rangle +  a^2(\eta) \left\langle \frac{\partial \phi_{s}}{\partial \vec{x}} \cdot \frac{\partial f_{s}}{\partial \vec{q}} \right \rangle \,,
\end{equation}
and subtracting Eq.~(\ref{eq:avgvlasov}) from Eq.~(\ref{eq:vlasovlong2}) leads to
\begin{equation}
\label{eq:flucvlasov}
	\frac{\partial \delta f_{l}}{\partial \eta} + \vec{q} \cdot \frac{\partial \delta f_{l}}{\partial \vec{x}} = a^2(\eta) \left[ \normOrd{\frac{\partial \phi_{l}}{\partial \vec{x}} \cdot \frac{\partial f_{l}}{\partial \vec{q}}} +\frac{\partial}{\partial \delta f_{l}}  	\left\langle \left[\frac{\partial \phi_{s}}{\partial \vec{x}} \cdot \frac{\partial f_{s}}{\partial \vec{q}} \right]_{\Lambda} \right\rangle_{\delta f_{l}} \Bigg|_{\delta f_{l}=0} \delta f_{l}(\eta,\vec{x},\vec{q}) + \cdots + \epsilon_{\Lambda} \right] \,.
\end{equation}
Let us reinforce that since $f_{l} = \bar{f}_{l}+\delta f_{l}$ appears on the right-hand side of Eqs.~(\ref{eq:avgvlasov}) and (\ref{eq:flucvlasov}), these are two coupled set of equations. Note that Eq.~(\ref{eq:flucvlasov}) contains additional terms when compared to Eq.~(\ref{eq:vlasovfluc}). They parametrize the dependence of short-wavelength fluctuations on the presence of a long mode and the effects of stochasticity, and were completely ignored in Sec.~\ref{sec:ptbe}. From this we can already see that the framework developed in Sec.~\ref{sec:ptbe} is necessarily incomplete, but we will come back to this point in Sec.~\ref{sec:decoupling}. However, in Eq.~(\ref{eq:avgvlasov}) short-scale fluctuations act as an additional source to the background distribution function $\bar{f}_{l}(\eta,q)$, which was exactly the basis for including a nonzero average velocity dispersion in Sec.~\ref{sec:resbdf}, and naturally arises in the EFT framework. 

Let us recall from Sec.~\ref{sec:ptbe} that at zeroth order in the perturbative expansion $\bar{f}_{l}^{(0)}(\eta,q) \propto \delta^{(3)}(\vec{q})$ for cold dark matter, and $\delta f_{l}^{(0)}(\eta,\vec{k},\vec{q}) = 0$. Also, it is only at second order that the background distribution function picks up corrections (the first order contribution vanishes), in which case we can split as before
\begin{equation}
\label{eq:spliteft}
	\bar{f}_{l}^{(2)}(\eta,q) = \bar{f}_{l,\textrm{P}}^{(2)}(\eta,q) + \bar{f}_{l,\textrm{ctr}}^{(2)}(\eta,q) \,,
\end{equation}
where the perturbative piece satisfies the equation we solved in Sec.~\ref{sec:pt} 
\begin{equation}
\label{eq:backpertur}
		\frac{\partial \bar{f}_{l,\textrm{P}}^{(2)}}{\partial \eta} = a^2(\eta) \left\langle \frac{\partial \phi_{l}^{(1)}}{\partial \vec{x}} \cdot \frac{\partial \delta f_{l}^{(1)}}{\partial \vec{q}} \right \rangle \,,
\end{equation}
and the counterterm piece is sourced by short-scale fluctuations
\begin{equation}
	\label{eq:backresumed}
	\frac{\partial \bar{f}_{l,\textrm{ctr}}^{(2)}}{\partial \eta} = a^2(\eta) \left\langle \frac{\partial \phi_{s}}{\partial \vec{x}} \cdot \frac{\partial f_{s}}{\partial \vec{q}} \right \rangle \,,
\end{equation}
where we treat the right-hand side in Eq.~(\ref{eq:backresumed}) as an external source with leading contribution to second order in the expansion (see \footref{footnote1} for additional comments on this). We can then immediately write
\begin{equation}
\label{eq:eftback}
	\bar{f}_{l,\textrm{ctr}}^{(2)}(\eta,q;\Lambda) = \int_{0}^{\eta} d\eta' a^2(\eta') \left\langle \frac{\partial \phi_{s}(\eta',\vec{x})}{\partial \vec{x}} \cdot \frac{\partial f_{s}(\eta',\vec{x},\vec{q})}{\partial \vec{q}} \right \rangle \,,
\end{equation} 
where here we include explicitly the cutoff dependence of the counterterm contribution to the long-wavelength background distribution function, which will be omitted in what follows to simplify the notation. Note that Eq.~(\ref{eq:eftback}) is a total derivative with respect to momentum, and hence it does not renormalize the background (mass) density, according to Eq.~(\ref{eq:energydensity}). We previously used this result without proof in Eq.~(\ref{eq:taylor2}), and it also holds true for the perturbative piece of the background distribution function as we showed in Eq.~(\ref{eq:backrenor}).

We are now ready to obtain an expression for the average velocity dispersion squared in terms of short scale fluctuations. It will be convenient to first rephrase it as a short scale kinetic energy per unit mass, $\kappa(\eta)$, where [essentially repeating Eq.~(\ref{eq:resvelocity})]
\begin{equation}
\label{eq:resvelocityrepeat}
	\bar{\rho}(\eta) \kappa(\eta) \equiv \frac{1}{2} \bar{\rho}(\eta) \sigma^2_{\textrm{dis}}(\eta) = \frac{1}{2} a^{-5}(\eta) \int \frac{d^3 \vec{q}}{(2\pi)^3} q^2 \bar{f}_{l,\textrm{ctr}}^{(2)}(\eta,q) \,.
\end{equation}
This can be obtained from Eq.~(\ref{eq:eftback}) after multiplying it by $q^2$ followed by an integration over momentum. A detailed derivation can be found in Appendix \ref{sec:app3}, but the final result is
\begin{equation}
\label{eq:cosmic_energy_int}
	\kappa(\eta) = -a^{-2}(\eta) \int_{0}^{\eta} d\eta' a(\eta') \frac{d}{d\eta'}\left[a(\eta')u(\eta')\right] \,,
\end{equation}
where
\begin{equation}
\label{eq:pot}
	\bar{\rho}(\eta)u(\eta) = \frac{1}{2} \langle \phi_{s}(\eta,\vec{x}) \rho_{s}(\eta,\vec{x})\rangle \,,
\end{equation}
defines the short scale gravitational binding energy per unit mass, $u(\eta)$. A more familiar form of Eq.~(\ref{eq:cosmic_energy_int}) can be obtained by differentiating it with respect to superconformal time
\begin{equation}
\label{eq:cosmic_energy}
	\frac{d}{d\eta} (\kappa + u) + \mathcal{H} (2\kappa + u) = 0 \,,
\end{equation}
where we recall that $\mathcal{H} = d\log a/d\eta = a^2H$. This is nothing but the Layzer-Irvine equation \cite{peebles1980large}, which generalizes the notion of energy conservation to an expanding background \footnote{The expansion breaks time-translation symmetry, so energy is not conserved in general. From Eq.~(\ref{eq:cosmic_energy}) we see that energy is conserved only for virialized scales, for which $2\kappa+u=0$.}.

Let us also emphasize that both $\kappa$ and $u$ appearing in Eq.~(\ref{eq:cosmic_energy}) are short-scale quantities as defined in Eqs.~(\ref{eq:resvelocityrepeat}) and (\ref{eq:pot}), while the Layzer-Irvine equation is often phrased in terms of the total kinetic and potential energies with contributions from all scales. We present a derivation of Eq.~(\ref{eq:cosmic_energy_int}), and hence of Eq.~(\ref{eq:cosmic_energy}), from the Boltzmann equation in Appendix \ref{sec:app1}, which holds true because contributions from long-wavelengths separately satisfy the very same equation, and can hence be subtracted.

The integral form of the cosmic energy Eq.~(\ref{eq:cosmic_energy_int}) shows that a nonzero average velocity dispersion is effectively sourced by a short-scale gravitational binding energy, which is not accounted for in perturbation theory, and in fact should contribute to the counterterm as we will now argue in Sec.~\ref{sec:decoupling}.  

\subsection{The inevitableness of EFT methods}
\label{sec:decoupling}

In Sec.~\ref{sec:resbdf} we have seen that an EFT-like counterterm, Eq.~(\ref{eq:powerres}), naturally arises within an old-fashioned cosmological perturbation theory approach based directly on a perturbative expansion for the distribution function fluctuations in full phase space. In that framework the counterterm is sourced by a nonzero average velocity dispersion, which is itself linked to the short-scale gravitational binding energy via the Layzer-Irvine Eq.~(\ref{eq:cosmic_energy_int}) as we showed in Sec.~\ref{sec:eft}. 

We will now argue that this short-scale gravitational binding energy should directly contribute to the counterterm as well to ensure the self-consistency of the theory. This implies that the framework developed in Sec.~\ref{sec:resbdf} is not self-consistent since only the velocity dispersion directly contributes to the counterterm, although it is formally identical to the EFT as long as one takes the counterterm to be a free parameter. Another way of phrasing this is to say that the value obtained for the counterterm $c_{\textrm{ctr}}^{(3)}(a)$, from directly matching Eq.(\ref{eq:powerres}) to simulations or observations, would not agree with the one calculated via Eqs.~(\ref{eq:source-re}) and (\ref{eq:analyticsolagainapp}), with an average velocity dispersion squared $\sigma_{\textrm{dis}}^2(a)$ equally extracted from simulations or observations.

The statement that the short-scale gravitational binding energy should also contribute to the counterterm can be seen from the well-understood decoupling of virialized scales \cite{Peebles:2009hw}. For convenience, let us repeat here the Euler Eq.~(\ref{eq:euler})
\begin{equation}
	\label{eq:euleragain} 
	\Pi_{i}' + 4\mathcal{H} \Pi_{i} +2a \partial^{j} K_{ij} +a\rho \partial_{i} \phi = 0 \,,
\end{equation}
where
\begin{equation}
	\label{eq:stressagain}
	K_{ij} =  \frac{1}{2} a^{-5} \int \frac{d^3 \vec{q}}{(2\pi)^{3}} \, q_{i}q_{j} \, f \,,
\end{equation}
is the kinetic energy density tensor. But we can write from the Poisson Eq.~(\ref{eq:poisson}),
\begin{equation}
	\label{eq:algebra}
	\rho \partial_{i} \phi = (\rho - \bar{\rho}) \partial_{i} \phi + \bar{\rho} \partial_{i} \phi = \frac{\nabla^2 \phi \partial_{i} \phi }{4\pi G a^2} + \bar{\rho} \partial_{i} \phi = \partial^{j} W_{ij} + \bar{\rho} \partial_{i} \phi \,,
\end{equation}
where
\begin{equation}
	\label{eq:bindingtensor}
	W_{ij} = \frac{1}{4\pi G a^2} \left[\partial_{i} \phi \partial_{j} \phi - \frac{1}{2} \delta_{ij} (\vec{\nabla} \phi)^2 \right] \,,
\end{equation}
is the potential energy density tensor. The Euler Eq.~(\ref{eq:euleragain}) can then be written in the following form
\begin{equation}
	\label{eq:euleronceagain}
	\Pi_{i}' + 4\mathcal{H} \Pi_{i} + a \bar{\rho} \partial_{i} \phi + a\partial^{j} (2K_{ij}+W_{ij}) = 0 \,.
\end{equation}
Note that $K = \langle K^{i}_{i} \rangle$ is just the kinetic energy density, and similarly from Eq.~(\ref{eq:bindingtensor}), $W = \langle W^{i}_{i} \rangle = \langle \phi \rho \rangle/2$  
is the gravitational binding energy density, where we integrate by parts inside the spatial average \footnote{The ensemble average can also be thought of as a volume average, which involves integrating over position. This allows an integration by parts under the average sign. \label{footnote2}}, use Eq.~(\ref{eq:poisson}) once again and assume $\langle \phi \rangle=0$. For virialized scales, $2K + W=0$ and all nonlinear terms in the Euler Eq.~(\ref{eq:euleronceagain}) vanish exactly. Of course this simplified argument ignores the tensor structure of these quantities, but a full tensorial virial decoupling theorem can be derived from the collisionless Boltzmann equation  (see \cite{f2381f32-aadb-3c8a-a473-53334cac15e7} for example). It is important to emphasize that the decoupling of virialized scales is exact \cite{Baumann:2010tm}. This contrasts with, being significantly more constraining than, the expectation that contributions from virialized scales are parametrically suppressed but not exactly vanishing.

It is now straightforward to argue that the old-fashioned cosmological perturbation theory framework developed in Sec.~\ref{sec:resbdf} is necessarily incomplete, since the average velocity dispersion in Eq.~(\ref{eq:stressavgagain}) is the quantity that sources the counterterm $c^{(3)}_{\textrm{ctr}}(a)$ [see Eq.(\ref{eq:powerres})] via Eqs.(\ref{eq:source-re}) and (\ref{eq:analyticsolagainapp}). However, we also know that virial velocities within halos \cite{Bryan_1998,  Sheth:2000ii} give a significant (if not dominant) contribution to the average velocity dispersion, and hence to the counterterm as well. This is inconsistent with the decoupling of virialized scales.

The resolution to this paradox comes from the realization that the short scale gravitational binding energy should also directly contribute to the counterterm, in such a way that it vanishes whenever $2K + W=0$. Additionally, just averages of the short-scale kinetic and potential energies are not sufficient to accurately model the sound speed counterterm. Instead, one needs to consider how such averages respond to the presence of a long-wavelength fluctuation \cite{Nascimento:2024rbv, Baumann:2010tm, Carrasco:2012cv}. We can explicitly identify the origin of the deficiencies in the framework developed in Sections \ref{sec:pt} and \ref{sec:resbdf} by comparing Eqs.(\ref{eq:vlasovfluc}) and (\ref{eq:flucvlasov}) . The latter equation includes additional terms that are neglected in the former, which parametrize stochasticity and the response of short-wavelength fluctuations to the presence of a long mode. This shows that short scale nonlinearities directly backreact into the background and fluctuations to the distribution function alike, and it is not self-consistent to only model the backreactions to the background distribution function. We can overcome these deficiencies by keeping the effective sound speed counterterm as a free parameter in the model, to be fitted by observations or full cosmological simulations. In that sense, EFT methods emerge as a necessary framework to self-consistently model a non-zero velocity dispersion.

All of that being said, one way in which one may be able to improve on the standard EFTofLSS framework is to solve Eq.~(\ref{eq:flucvlasov}) perturbatively for the distribution function fluctuations while assuming knowledge of the fully nonlinear background distribution function, which can be measured from simulations. This is equivalent to resumming the effects of the background distribution function backreacting into its fluctuations to all orders in perturbation theory, while working with an EFT perturbative expansion for the fluctuations. In practice, this would look a lot like the standard EFTofLSS framework augmented with modified perturbation theory kernels. A similar idea was proposed recently in \cite{Garny:2025ovs}.
\section{Conclusion}
\label{sec:concl}

Effective field theory methods for large scale structure significantly improve on Standard Perturbation Theory (SPT) techniques by modeling deviations from an ideal fluid \cite{Baumann:2010tm, Carrasco:2012cv, Carroll:2013oxa, Pietroni:2011iz, Carrasco:2013mua, Porto:2013qua, Vlah:2015sea, Zaldarriaga:2015jrj, Baldauf:2016sjb, Braganca:2023pcp, Nishimichi:2020tvu}, and are now ubiquitous in analysis pipelines of large scale structure surveys \cite{Euclid:2023tog,Chudaykin:2020aoj, Ivanov:2019pdj,Philcox:2022frc, Taule:2023izt, Beutler:2011hx, Blake:2011rj, Chudaykin:2022nru, Taule:2024bot, Aviles:2024zlw, Colas:2019ret, Simon:2022csv, DAmico:2022osl, DESI:2024jis, Chen:2024vvk, Chen:2022jzq, Moretti:2023drg}. 

In its simplest form \footnote{Modeling the power spectrum of the matter field, in real space (as opposed to redshift space), at one-loop in perturbation theory. This is assumed in the discussion that follows.}, the EFTofLSS comes at the cost of adding one free nuisance parameter, the effective sound speed. This is arguably not a desirable feature, after all N-body simulations have no free parameters, and we are entering a new era where efficient emulators are available to interpolate the predictions from simulations in broad regions of parameter (and even theory) space \cite{Euclid:2020rfv, Angulo:2020vky, Heitmann:2013bra, Nishimichi:2018etk, DeRose:2018xdj, Wibking:2017slg, Winther:2019mus, Brando:2022gvg, Mauland:2023pjt, Jamieson:2024fsp, Jamieson:2022lqc}. This naturally raises the question of whether or not the additional free parameter is really necessary (and this becomes even more relevant in light of recent studies on prior volume effects in the EFTofLSS \cite{Ivanov:2019pdj, Philcox:2021kcw, Ivanov:2023qzb, Holm:2023laa, Simon:2022lde, Maus:2024sbb, Carrilho:2022mon, Donald-McCann:2023kpx}).

The starting point of Standard Perturbation Theory (SPT) methods for large scale structure is the assumption of negligible stress tensor \cite{Bernardeau:2001qr}. This leaves open the possibility that a perturbative approach can account for these effects by avoiding to truncate the Boltzmann hierarchy at the level of the Euler equation \cite{Garny:2022tlk, Garny:2022kbk, Garny:2025ovs, Erschfeld:2018zqg, Dominguez:2000dt,mcdonald2011generate, Kitaura:2023mfc, Matarrese:2007wc, McDonald:2017ths, Aviles:2015osc, tassev2011helmholtz, Valageas:2003gm}. In this work we revisit this issue in light of a framework for large scale structure perturbation theory, that directly solves for the distribution function in full phase space, expanding upon earlier work \cite{Valageas:2001df}. This approach circumvents the need to artificially truncate the Boltzmann hierarchy and hence relaxes the usual \textit{apriori} assumption of a negligible velocity dispersion.

In Sec.~\ref{sec:ptbe} we introduce the framework underlying the cosmological perturbation theory in phase space. It is based on the coupled set of Boltzmann Eqs.~(\ref{eq:vlasovavg}) and (\ref{eq:vlasovfluc}) for the (ensemble) average distribution function and its fluctuations, respectively. We show that perturbatively solving this coupled set of equations directly in phase space reproduces the familiar SPT kernels, which underscores the statement that a negligible velocity dispersion and vorticity should be viewed as a consequence of the perturbative expansion, rather than an assumption \cite{Valageas:2001df}. 

We begin Sec.~\ref{sec:eftoflss} by showing that a nonzero average velocity dispersion can be easily incorporated in the framework by perturbatively solving for the fluctuations in the distribution function, while assuming knowledge of its fully nonlinear ensemble average. This leads to a new contribution to the one-loop power spectrum with the exact same form as the effective sound speed counterterm in the EFTofLSS [see Eq.~(\ref{eq:powerres})]. This result corroborates similar previous findings in the literature \cite{Garny:2022kbk,Aviles:2015osc, McDonald:2017ths}. We then proceed to make a direct connection to EFT methods, and argue that our framework is necessarily incomplete because it misses contributions to the counterterm from the short-scale gravitational binding energy, and the response of short-wavelength fluctuations to the presence of a long mode. EFT methods then arise as an inevitable framework to self-consistently model a nonzero velocity dispersion. This is a practical example of the importance in having theoretical control over short distance fluctuations, in order to write a sensible theory. 

Beyond the results summarized above, we intend this manuscript to be used as a useful pedagogical reference for numerical calculations in SPT with the full time-dependence of $\Lambda$CDM kernels, up to third order in the perturbative expansion. For that purpose, simple analytic formulas for the calculation of time-dependent coefficients can be found in Appendix \ref{sec:app1}. We also include explicit formulas for the one-loop power spectrum and tree-level bispectrum in Appendices \ref{sec:app4} and \ref{sec:app5}, respectively. 

To conclude, we land at a picture of the microphysics behind the EFTofLSS that suggests a close analogy to chiral perturbartion theory in QCD \cite{Gasser:1983yg}: The stress tensor acts like an order parameter (analogous to the quark condensate) which vanishes in the perturbative regime, but picks up a nonzero value in the strongly coupled regime which follows after gravitational collapse (corresponding to the QCD phase transition). Moreover, the triaxial nature of gravitational collapse implies that an approximate symmetry of the perturbative regime, i.e. isotropy, is spontaneously broken as gravitational collapse first happens along a given axis. This is analogous to the breakdown of the approximate chiral symmetry in QCD. Furthermore, just as in QCD lattice simulations \cite{Wilson:1974sk} are available to compute nonperturbative processes of interest without the need to introduce any additional free parameters, the same is true in LSS with N-body simulations. Nonetheless, in the same way that chiral perturbation theory has proved itself to be a very valuable tool in QCD, recent progress over the past decade or so has established the EFTofLSS as a great framework to interpret both simulated and real data pertaining the evolution of LSS within an analytic perturbative framework. In this context, the additional free parameter is simply the price one needs to pay in order to parametrize intrinsically nonperturbative effects within a perturbative framework.

\acknowledgments

This work is supported by the Department of Energy grants DE-SC0023183 and DE-SC0011637 and the Dr. Ann Nelson Endowed Professorship. We thank Roman Scoccimarro and Mathias Garny for excellent comments and discussions. CN thanks Matthew McQuinn for many helpful conversations and Patrick McDonald for fruitful conversation and for pointing out relevant references. All the numerical calculations and plots in this paper were made with Mathematica \cite{Mathematica}.
\appendix 

\section{Analytic solution to time-dependent coefficients}
\label{sec:app1}

In this section we derive analytic formulas for the time-dependent coefficients in SPT, entering Eqs.~(\ref{eq:separable-2nd}) and (\ref{eq:separable-3rd}), within the full $\Lambda$CDM cosmology. To accomplish this, we will first need to find an analytic solution to the differential equation
\begin{equation}
\label{eq:sourced}
	\frac{d^2c}{da^2} +\frac{1}{a} \left(3+\frac{d\log H}{d\log a}\right) \frac{dc}{da} - \frac{3}{2} \Omega_{\textrm{m},0} H_{0}^{2} \frac{c(a)}{a^5H(a)^2} = s(a) \,,
\end{equation} 
for the coefficient $c(a)$, given some source function $s(a)$. To achieve this, we apply the same trick that works in the $s(a)=0$ case. The first step is to note that the Hubble expansion rate $H(a)$ satisfies the homogeneous equation in $\Lambda \textrm{CDM}$ (one can even add a nonzero curvature), that is:
\begin{equation}
\label{eq:hubble}
	\frac{d^2H}{da^2} +\frac{1}{a} \left(3+\frac{d\log H}{d\log a}\right) \frac{dH}{da} - \frac{3}{2} \Omega_{\textrm{m},0} H_{0}^{2} \frac{1}{a^5H(a)} = 0 \,.
\end{equation}

To take advantage of this, we define a new function $g(a)=c(a)/H(a)$, and combine Eqs.(\ref{eq:sourced}) and (\ref{eq:hubble}) to arrive at
\begin{equation}
\label{eq:stepaux}
	\frac{d}{da} \left[\frac{dg}{da} a^3 H^3(a) \right] = a^3H^2(a) s(a) \,,
\end{equation}
and the solution to this can be simply obtained by integrating the equation twice with respect to the scale factor, which reads in terms of $c(a)=H(a)g(a)$
\begin{equation}
\begin{split}
\label{eq:finalaux}
	c(a) = & H(a) \Bigg[ g(a_{i}) +g'(a_{i}) a_{i}^3 H^{3}(a_{i}) \int_{a_{i}}^{a} \frac{da'}{(a')^{3} H^{3}(a')} \  + \\ & + \  \int_{a_{i}}^{a} \frac{da'}{(a')^3 H^{3}(a')} \int_{a_{i}}^{a'} da'' (a'')^{3}H^{2}(a'') s(a'') \Bigg] \,.
\end{split}
\end{equation} 
In the case of $s(a)=0$, Eq.(\ref{eq:sourced}) reduces to Eq.(\ref{eq:ie-lo-ode-time}) satisfied by the linear growth factor. According to Eq.(\ref{eq:finalaux}), the general solution is then a linear combination of the growing and decaying modes, $c_{+}(a)$ and $c_{-}(a)$ respectively, where
\begin{equation}
\label{eq:generalsol}
\begin{split}
	& c_{+}(a) = H(a) \int_{0}^{a} \frac{da'}{(a')^{3} H^{3}(a')}  \\ & c_{-}(a) = H(a) \,,
\end{split}
\end{equation}  
and we set $a_{i}=0$. If we drop the decaying mode and normalize the linear growth according to $D_{\textrm{L}}(a=1)=1$, we arrive at:
\begin{equation}
	\label{eq:analyticapp}
	D_{\textrm{L}}(a) = \frac{H(a)}{H_{0}} \left[\int_{0}^{1} \frac{da'}{(a')^3 H^{3}(a')}\right]^{-1} \int_{0}^{a} \frac{da'}{(a')^3 H^{3}(a')} \,,
\end{equation} 
hence reproducing a familiar result, which we used in Sec.~\ref{sec:ptbe}. However, in the more general case of a nonzero source the particular solution typically grows faster than the homogeneous solutions and eventually dominates. We then obtain
\begin{equation}
\label{eq:analyticsolagainapp}
	c(a) = H(a) \int_{0}^{a} \frac{da'}{(a')^3 H^{3}(a')} \int_{0}^{a'} da'' (a'')^{3} H^{2}(a'') s(a'') \,.
\end{equation}

With this basic ingredient we can write down formulas for the time-dependent coefficients in SPT. Starting at second order in the perturbative expansion, we show in Appendix \ref{sec:app2} that the coefficients $c^{(2)}_{i}(a)$ are solutions to the differential Eq.(\ref{eq:sourced}), with source functions
\begin{equation}
	\label{eq:sources-2nd}
	\begin{split}
		& s^{(2)}_{1} = \frac{D_{\textrm{L}}^2 f}{a^2} \left[2(1+f) + \frac{d\log H}{d\log a} + \frac{d\log f}{d\log a} \right] \\ & s^{(2)}_{2} = \frac{D_{\textrm{L}}^2 f^2}{a^2} \,,
	\end{split}
\end{equation}
where the dependence on scale factor is implicit in Eq.(\ref{eq:sources-2nd}). The solutions are then obtained from a direct application of Eq.~(\ref{eq:analyticsolagainapp}), which can be easily evaluated numerically. For an Einstein-de Sitter (EdS) universe with $\Omega_{\textrm{m}}(a)=1$, $D_{\textrm{L}}(a)=a$ which implies $f(a)=1$. In this case the source terms in Eq.(\ref{eq:sources-2nd}) simplify to $s^{(2)}_{1}(a) = 5/2$ and $s^{(2)}_{2}(a) = 1$, and it becomes straightforward to derive the time-dependent coefficients: $c^{(2)}_{1}(a) = 5a^2/7$ and $c^{(2)}_{2}(a) = 2a^2/7$. This motivates the familiar EdS approximation to time-dependent coefficients in the general $\Lambda$CDM cosmology: 
\begin{equation}
	\label{eq:eds_2nd}
	\begin{split}
		& c^{(2)}_{1, \textrm{EdS}}(a) \approx \frac{5}{7}D_{\textrm{L}}^2(a)\\ & c^{(2)}_{2, \textrm{EdS}}(a) \approx \frac{2}{7} D_{\textrm{L}}^2(a) \,.
	\end{split}
\end{equation}

\begin{figure}
	\centering
	\includegraphics[width=0.75\textwidth]{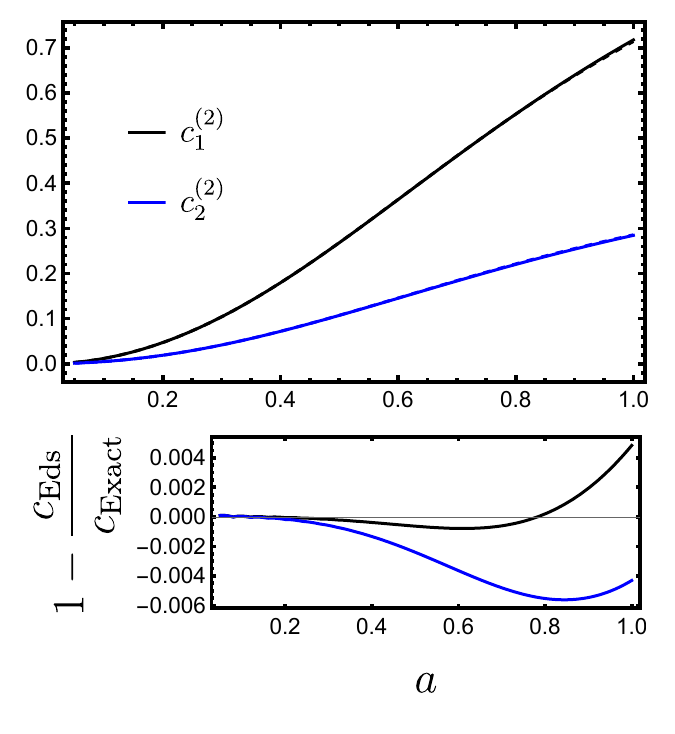}
	\caption{Time-dependent coefficients for the second order perturbation theory kernels, in our fiducial cosmology, as a function of the scale factor. Solid and dashed lines (almost indistinguishable) correspond to exact solutions from Eqs.~(\ref{eq:analyticsolagainapp}) and (\ref{eq:sources-2nd}), and the EdS approximation in Eq.~(\ref{eq:eds_2nd}), respectively. The lower plot shows the relative difference between the two, which is always at the sub-percent level.}
	\label{fig:2nd}
\end{figure}

In Fig.~\ref{fig:2nd} we plot these time-dependent coefficients in our fiducial cosmology. The exact solutions from Eqs.~(\ref{eq:analyticsolagainapp}) and (\ref{eq:sources-2nd}) are shown as solid lines, and the approximated ones from Eq.~(\ref{eq:eds_2nd}) as dashed lines. The EdS approximation works extremely well, with $<1\%$ errors. 

The third order time-dependent coefficients also satisfy the same differential Eq.(\ref{eq:sourced}), with new source functions 
\begin{equation}
	\label{eq:sources-3rd}
	\begin{split}
		& s^{(3)}_{1} = \frac{fD_{\textrm{L}}}{a}\frac{dc^{(2)}_{1}}{da} + \frac{fD_{\textrm{L}}}{a^2} \left(2+f+\frac{d\log H}{d\log a} + \frac{d\log f}{d\log a}\right)c^{(2)}_{1} \\ & s^{(3)}_{2} = \frac{fD_{\textrm{L}}}{a}\frac{dc^{(2)}_{2}}{da} + \frac{fD_{\textrm{L}}}{a^2} \left(2+f+\frac{d\log H}{d\log a} + \frac{d\log f}{d\log a}\right)c^{(2)}_{2} \\ & s^{(3)}_{3} = \frac{3}{2} \Omega_{\textrm{m},0} H_{0}^{2} \frac{D_{\textrm{L}}c^{(2)}_{1}}{a^5H^2} + \frac{fD_{\textrm{L}}}{a} \frac{dc^{(2)}_{1}}{da} - \frac{f^2 D_{\textrm{L}}^3}{a^2} \\ & s^{(3)}_{4} = \frac{3}{2} \Omega_{\textrm{m},0} H_{0}^{2} \frac{D_{\textrm{L}}c^{(2)}_{2}}{a^5H^2} + \frac{fD_{\textrm{L}}}{a} \frac{dc^{(2)}_{2}}{da} + \frac{ f^2 D_{\textrm{L}}^3}{a^2} \\ & s^{(3)}_{5} = 2 \frac{fD_{\textrm{L}}}{a}\frac{dc^{(2)}_{1}}{da} -2 \frac{f^2 D_{\textrm{L}}^3 }{a^2} \\ & s^{(3)}_{6} = 2 \frac{fD_{\textrm{L}}}{a}\frac{dc^{(2)}_{2}}{da} \,.
	\end{split}
\end{equation}
As before, solutions are obtained from a direct application of Eq.~(\ref{eq:analyticsolagainapp}). 
One can derive the EdS approximation to the third order coefficients in the exact same way as done in the previous case of second order coefficients. This reads
\begin{equation}
	\label{eq:eds_3rd}
	\begin{split}
		& c^{(3)}_{1, \textrm{EdS}}(a) \approx \frac{5}{18}D_{\textrm{L}}^{3}(a)\\ & c^{(3)}_{2, \textrm{EdS}}(a) \approx \frac{1}{9} D_{\textrm{L}}^{3}(a) \\ & c^{(3)}_{3, \textrm{EdS}}(a) \approx \frac{1}{6} D_{\textrm{L}}^{3}(a) \\ & c^{(3)}_{4, \textrm{EdS}}(a) \approx \frac{2}{9} D_{\textrm{L}}^{3}(a) \\ & c^{(3)}_{5, \textrm{EdS}}(a) \approx \frac{2}{21} D_{\textrm{L}}^{3}(a) \\ & c^{(3)}_{6, \textrm{EdS}}(a) \approx \frac{8}{63} D_{\textrm{L}}^{3}(a) \,.
	\end{split}
\end{equation}

\begin{figure}
	\centering
	\includegraphics[width=0.75\textwidth]{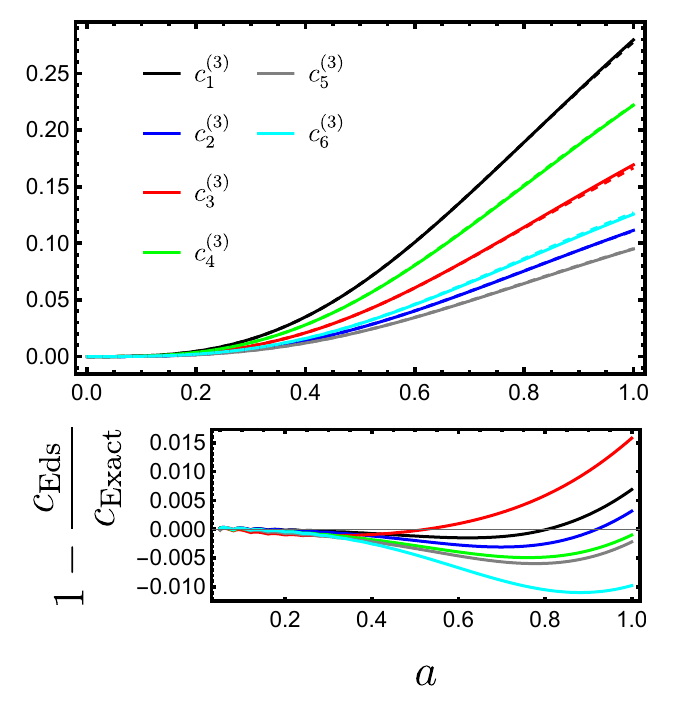}
	\caption{Time-dependent coefficients for the third order perturbation theory kernels, in our fiducial cosmology, as a function of the scale factor. Solid and dashed lines (almost indistinguishable) correspond to exact solutions from Eqs.~(\ref{eq:analyticsolagainapp}) and (\ref{eq:sources-3rd}), and the EdS approximation in Eq.~(\ref{eq:eds_3rd}), respectively. The lower plot shows the relative difference between the two, which is at the percent level. }
	\label{fig:3rd}
\end{figure}

In Fig.~\ref{fig:3rd} we plot the time-dependent coefficients for the third order perturbation theory kernels, $c^{(3)}_{i}(a)$ with $i=\overline{1,6}$. The exact solutions from Eqs.~(\ref{eq:analyticsolagainapp}) and (\ref{eq:sources-3rd}) are shown as solid lines, and the approximated ones from Eq.~(\ref{eq:eds_3rd}) as dashed lines. The EdS approximation still works quite well, but slightly worse than in the second order case shown in Fig.\ref{fig:2nd}, with $\lesssim1.5\%$ errors overall. 

The formulas presented here are fully equivalent to the many different approaches to computing time-dependent coefficients in SPT, that can be found on the extensive literature on this subject (see, e.g., \cite{Hartmeier:2023brx, Takahashi:2008yk, Bernardeau:1993qu, Fasiello:2022lff, Fasiello:2016qpn, Choustikov:2023uyk, Lewandowski:2016yce, Fujita:2020xtd, Donath:2020abv, Schmidt:2020ovm, Rampf:2022tpg, Garny:2022fsh}). Nevertheless, we find Eq.~(\ref{eq:analyticsolagainapp}) to be particularly simple and easy to implement in practice.

\section{Computing second order diagrams}
\label{sec:app2}

We will now compute all relevant diagrams at second order in perturbation theory, starting with the diagram in Fig.~\ref{fig:diagram5} for the background distribution function at second order in the perturbative expansion. It represents two insertions of $\phi^{(1)}$ into the second order iterative solution. To compute this, we first substitute the second line of Eq.~(\ref{eq:first}) into the first line of Eq.~(\ref{eq:second}). This yields
\begin{equation}
\label{eq:diagram5step1}
\begin{split}
	\bar{f}^{(\textrm{2nd})}(\eta,q) =& \int_{0}^{\eta} d\eta' a^{2}(\eta') \int_{0}^{\eta'} d\eta'' a^{2}(\eta'') \int_{\{\vec{k}_1, \vec{k}_2\}}^{\vec{k}} \langle \phi(\eta',\vec{k}_{1})   \phi(\eta'',\vec{k}_{2}) \rangle \ \times \\ & \times i\vec{k}_{1} \cdot \frac{\partial}{\partial \vec{q}} \left[ i\vec{k}_{2} \cdot \frac{\partial \bar{f}^{(0)}}{\partial \vec{q}} e^{-i\vec{k}_{2} \cdot \vec{q} (\eta'-\eta'')}  \right] \,.
\end{split}
\end{equation}
We now insert two copies of $\phi^{(1)}$, using Eqs.~(\ref{eq:poisson2}), (\ref{eq:separable}) and (\ref{eq:2pf}) to obtain
\begin{equation}
	\label{eq:backrenorapp}
	\begin{split}
		\bar{f}^{(2)}(\eta,q) = - &  \left(\frac{3}{2}\Omega_{\textrm{m},0}H_0^2\right)^2 \int_{0}^{\eta} d\eta' a(\eta') D_{\textrm{L}}(\eta') \int_{0}^{\eta'} d\eta'' a(\eta'') D_{\textrm{L}}(\eta'') \, \times \\ & \times \int \frac{d^3\vec{k}'}{(2\pi)^3} \frac{P_{\textrm{L}}(k')}{(k')^4} i\vec{k}' \cdot \frac{\partial}{\partial \vec{q}} \left[i\vec{k}' \cdot \frac{\partial \bar{f}^{(0)}}{\partial \vec{q}} e^{i\vec{k}' \cdot \vec{q} (\eta'-\eta'')} \right] \,.
	\end{split}
\end{equation}
In what follows we omit time dependencies for simplicity. We evaluate next
\begin{equation}
\label{eq:diagram5step2}
\begin{split}
	& a^{-5} \int \frac{d^3 \vec{q}}{(2\pi)^{3}} q^2 \bar{f}^{(2)}(q) = \\ &= - \left(\frac{3}{2}\Omega_{\textrm{m},0}H_0^2\right)^2 \int_{0}^{\eta} d\eta' a(\eta') D_{\textrm{L}}(\eta') \int_{0}^{\eta'} d\eta'' a(\eta'') D_{\textrm{L}}(\eta'') \  \times \\ & \times \ \int \frac{d^3\vec{k}'}{(2\pi)^3} \frac{P_{\textrm{L}}(k')}{(k')^4} \int \frac{d^3 \vec{q}}{(2\pi)^{3}} q^2 \ i\vec{k}' \cdot \frac{\partial}{\partial \vec{q}} \left[i\vec{k}' \cdot \frac{\partial \bar{f}^{(0)}}{\partial \vec{q}} e^{i\vec{k}' \cdot \vec{q} (\eta'-\eta'')} \right] \,.
\end{split}
\end{equation}
The integral over momentum can be simplified via integrating by parts twice, after which we can use $\bar{f}^{(0)}(q) \propto \delta^{(3)}(\vec{q})$ and Eq.~(\ref{eq:energydensity}) to arrive at
\begin{equation}
\label{eq:diagram5step3}
\begin{split}
	 & a^{-5} \int \frac{d^3 \vec{q}}{(2\pi)^{3}} q^2 \bar{f}^{(2)}(q) = \\ &= 2a^{-2} \bar{\rho} \int_{0}^{\infty} \frac{dk'}{2\pi^2} P_{\textrm{L}}(k') \times  \left(\frac{3}{2}\Omega_{\textrm{m},0}H_0^2\right)^2 \int_{0}^{\eta} d\eta' a(\eta') D_{\textrm{L}}(\eta') \int_{0}^{\eta'} d\eta'' a(\eta'') D_{\textrm{L}}(\eta'') \,.
	\end{split}
\end{equation}
The final step is to evaluate the time integrals. This is straightforward to do using Eq.~(\ref{ie-lo-ode}) since the integrands can be written as total derivatives, and the final result reads:
\begin{equation}
	\label{eq:stresscancelapp}
	a^{-5} \int \frac{d^3 \vec{q}}{(2\pi)^3} q^2 \bar{f}^{(2)}(q) = \bar{\rho} \, a^2 H^2 f^2 D_{\textrm{L}}^2 \int_{0}^{\infty} \frac{dk'}{2\pi^2} P_{\textrm{L}}(k') \,.
\end{equation}
This equation was used in the main text to show that the averaged stress tensor indeed vanishes as a result of the perturbative expansion. 

We next compute the two diagrams in Fig.~\ref{fig:diagram4}, for the distribution function fluctuation at second order in perturbation theory, to derive Eqs.~(\ref{eq:sourced}) and (\ref{eq:sources-2nd}). Denoting the diagram on the left of Fig.~\ref{fig:diagram4} by $\mathcal{D}_{1}$, and the diagram on the right by $\mathcal{D}_{2}$, we have:
\begin{equation}
\label{eq:summed_diagrams}
	\delta f^{(2)} = \mathcal{D}_{1} + \mathcal{D}_{2} \,.
\end{equation}

We start with $\mathcal{D}_{1}$, where we need to insert one $\phi^{(2)}$ into the first order iterative solution, i.e., the second line in Eq.~(\ref{eq:first}). This reads using Eq.~(\ref{eq:poisson2}),
\begin{equation}
\label{eq:diagram4step1}
	\mathcal{D}_{1} = - \frac{3}{2} \Omega_{\textrm{m,0}}H_{0}^2 \int_{0}^{\eta} d\eta' a^2(\eta') e^{-i \vec{k} \cdot \vec{q} (\eta-\eta')} \delta^{(2)}(\eta',\vec{k}) \frac{i\vec{k}}{k^2} \cdot \frac{\partial \bar{f}^{(0)}}{\partial \vec{q}} \,.
\end{equation} 
For $\mathcal{D}_{2}$, we need two insertions of $\phi^{(1)}$ into the second order iterative solution for the distribution function fluctuation, i.e., second line in Eq.~(\ref{eq:second}). Combining it with Eqs.~(\ref{eq:first}), (\ref{eq:poisson2}) and (\ref{eq:separable}) we obtain
\begin{equation}
	\label{eq:diagram4step2}
	\begin{split}
	& \mathcal{D}_{2} = \left(\frac{3}{2}\Omega_{\textrm{m},0}H_0^2\right)^2 \int_{0}^{\eta} d\eta' a(\eta') D_{\textrm{L}}(\eta') \int_{0}^{\eta'} d\eta'' a(\eta'') D_{\textrm{L}}(\eta'') \, \times \\ & \times \int_{\{\vec{k}_1, \vec{k}_2\}}^{\vec{k}}  \frac{1}{k_{1}^2 k_{2}^2} \normOrd{\delta_{\textrm{L}}(\vec{k}_{1})\delta_{\textrm{L}}(\vec{k}_{2})} e^{-i\vec{k} \cdot \vec{q}(\eta-\eta')} i\vec{k}_{1} \cdot \frac{\partial}{\partial \vec{q}} \left[i\vec{k}_{2} \cdot \frac{\partial \bar{f}^{(0)}}{\partial \vec{q}} e^{-i\vec{k}_{2} \cdot \vec{q} (\eta'-\eta'')} \right] \,.
	\end{split}
\end{equation}
To arrive at an integral equation for the density contrast, we need to integrate over momentum according to Eq.~(\ref{eq:energydensity}). That is,
\begin{equation}
\label{eq:diagram4step3}
	\delta^{(2)} = \hat{\mathcal{D}}_{1} + \hat{\mathcal{D}}_{2} \,,
\end{equation}
where
\begin{equation}
\label{eq:diagram4step4}
	\hat{\mathcal{D}}_{i} = \frac{1}{a^3 \bar{\rho}} \int \frac{d^3 \vec{q}}{(2\pi)^3} \mathcal{D}_{i} \,.
\end{equation}
Since $\bar{f}^{(0)}(q) \propto \delta^{(3)}(\vec{q})$, the momentum integrals can be easily evaluated upon integrating by parts a few times to obtain
\begin{equation}
\label{eq:diagram4step5}
	\hat{\mathcal{D}}_{1} = \frac{3}{2}\Omega_{\textrm{m},0}H_0^2 \int_{0}^{\eta} d\eta' a(\eta') (\eta-\eta') \delta^{(2)}(\eta',\vec{k}) \,,
\end{equation} 
and
\begin{equation}
\label{eq:diagram4step6}
\begin{split}
	& \hat{\mathcal{D}}_{2} = \left(\frac{3}{2}\Omega_{\textrm{m},0}H_0^2\right)^2 \int_{0}^{\eta} d\eta' a(\eta') D_{\textrm{L}}(\eta')(\eta - \eta') \int_{0}^{\eta'} d\eta'' a(\eta'') D_{\textrm{L}}(\eta'')(\eta - \eta'') \times h_{1}^{(2)}(\vec{k}) \\ & + \left(\frac{3}{2}\Omega_{\textrm{m},0}H_0^2\right)^2 \int_{0}^{\eta} d\eta' a(\eta') D_{\textrm{L}}(\eta')(\eta - \eta')^2 \int_{0}^{\eta'} d\eta'' a(\eta'') D_{\textrm{L}}(\eta'') \times h_{2}^{(2)}(\vec{k}) \small \,,
\end{split}
\end{equation}
where the scale dependent functions $h_{i}^{(2)}(\vec{k})$ are defined in Eq.~(\ref{eq:scale}). The time integrals in Eq.~(\ref{eq:diagram4step6}) can be greatly simplified by using Eq.~(\ref{ie-lo-ode}), followed by a few integration by parts. The final result is
\begin{equation}
\label{eq:diagram4step7}
	\hat{\mathcal{D}}_{2} = \frac{1}{2} D_{\textrm{L}}^{2}(\eta) \ h_{1}^{(2)}(\vec{k}) + \int_{0}^{\eta} d\eta' \left(\frac{dD_{\textrm{L}}}{d\eta'}\right)^{2} (\eta-\eta') \ h_{2}^{(2)}(\vec{k}) \,.
\end{equation}
We are now ready to combine Eqs.~(\ref{eq:diagram4step3}), (\ref{eq:diagram4step5}) and (\ref{eq:diagram4step7}) to write down the integral equation satisfied by the second order density contrast:
\begin{equation}
\label{eq:diagram4step8}
\begin{split}
	\delta^{(2)}(\eta,\vec{k}) &= \frac{3}{2}\Omega_{\textrm{m},0}H_0^2 \int_{0}^{\eta} d\eta' a(\eta') (\eta-\eta') \delta^{(2)}(\eta',\vec{k}) +  \frac{1}{2} D_{\textrm{L}}^{2}(\eta) \ h_{1}^{(2)}(\vec{k}) \\ & + \int_{0}^{\eta} d\eta' \left(\frac{dD_{\textrm{L}}}{d\eta'}\right)^{2} (\eta-\eta') \ h_{2}^{(2)}(\vec{k}) \,.
\end{split}
\end{equation}
This can be mapped into a second order differential equation by taking two derivatives with respect to superconformal time,
\begin{equation}
\label{eq:diff2nd}
	\frac{d^2 \delta^{(2)}}{d\eta^2} - \frac{3}{2}\Omega_{\textrm{m},0}H_0^2 a(\eta) \delta^{(2)}(\eta,\vec{k}) = \frac{1}{2} \frac{d^2}{d\eta^2} D_{\textrm{L}}^2(\eta) \ h_{1}^{(2)}(\vec{k}) + \left(\frac{dD_{\textrm{L}}}{d\eta}\right)^2 \ h_{2}^{(2)}(\vec{k}) \,.
\end{equation}
The final step is to change the time variable from superconformal time $\eta$ to the scale factor $a$, to arrive at the differential Eq.(\ref{eq:sourced}). In terms of the separable ansatz of Eq.~(\ref{eq:separable-2nd}), the source terms are found to be
\begin{equation}
\label{eq:sources2nd}
\begin{split}
& s_{1}^{(2)} = \frac{1}{a^6H^2}  \ \frac{1}{2} \frac{d^2}{d\eta^2} D_{\textrm{L}}^2 = \frac{D_{\textrm{L}}^2 f}{a^2} \left[2(1+f) + \frac{d\log H}{d\log a} + \frac{d\log f}{d\log a} \right] \\ & s_{2}^{(2)} = \frac{1}{a^6H^2} \left(\frac{dD_{\textrm{L}}}{d\eta}\right)^2 = \frac{D_{\textrm{L}}^2 f^2}{a^2} \,.
\end{split}
\end{equation}

The steps involved in this calculation can be summarized as follows: First integrate over momentum to obtain $\hat{\mathcal{D}}$, according to Eq.~(\ref{eq:diagram4step4}), from a given diagram $\mathcal{D}$. Then differentiate it twice with respect to superconformal time followed by a division by $a^{6}H^{2}$ to obtain the source term upon factoring out the scale dependent piece. This prescription was used in the main text to derive Eq.~(\ref{eq:source-re}). However, note that this recipe cannot be applied to diagrams such as the leftmost ones in Figs.~\ref{fig:diagram4} and \ref{fig:diagram6}, i.e. with all wiggly lines attached to a single circle, because they contribute to the homogeneous part of the differential equation [for example the left-hand side of Eq.~(\ref{eq:diff2nd})], and do not appear as source terms.

We omit explicit calculations of the third order diagrams in Fig.~\ref{fig:diagram6} since they rely on the exact same tools, but are more tedious to write down as one would expect from a higher order calculation. There are two basic tricks one has to keep in mind, which appeared already at the second order level. First, to integrate by parts over momentum to leverage the fact that $\bar{f}^{(0)}(q) \propto \delta^{(3)}(\vec{q})$. Second, to simplify the integrals over superconformal time by using Eq.~(\ref{ie-lo-ode}) followed by a few integration by parts.
 
\section{Layzer-Irvine equation}
\label{sec:app3}

Here we present a derivation of the cosmic energy Eq.~(\ref{eq:cosmic_energy_int}). Our starting point is Eq.~(\ref{eq:eftback}) for the counterterm contribution to the background distribution function, sourced by short-wavelength fluctuations,  
\begin{equation}
	\label{eq:eftbackapp}
	\bar{f}_{l,\textrm{ctr}}(\eta,q) = \int_{0}^{\eta} d\eta' a^2(\eta') \left\langle \frac{\partial \phi_{s}(\eta',\vec{x})}{\partial \vec{x}} \cdot \frac{\partial f_{s}(\eta',\vec{x},\vec{q})}{\partial \vec{q}} \right \rangle \,,
\end{equation} 
The first step is to multiply Eq.~(\ref{eq:eftbackapp}) by $q^2$ and integrate over momentum
\begin{equation}
	\label{eq:cosmic_energy_step1}
	\int \frac{d^3 \vec{q}}{(2\pi)^3} q^2 \bar{f}_{l,\textrm{ctr}}^{(2)}(\eta,q) = -2\int_{0}^{\eta} d\eta' a^{6}(\eta') \left\langle \frac{\partial \phi_{s}}{\partial \vec{x}}\Big|_{\eta'} \cdot \vec{\Pi}_{s}(\eta',\vec{x}) \right\rangle \,,
\end{equation}
where we integrated by parts once and used Eq.~(\ref{eq:momentum}). Next we integrate by parts inside the spatial average (see \footref{footnote2}), and use Eqs.~(\ref{eq:continuity}) and (\ref{eq:poisson}) to obtain \footnote{All equations which are linear in the distribution function are satisfied separately by both long and short-wavelength pieces. }
\begin{equation}
	\label{eq:cosmic_energy_step2}
	\int \frac{d^3 \vec{q}}{(2\pi)^3} q^2 \bar{f}_{l,\textrm{ctr}}^{(2)}(\eta,q) = -2\int_{0}^{\eta} d\eta' a^{2}(\eta') \left\langle \phi_{s}(\eta',\vec{x}) \frac{\partial}{\partial \eta} \left[a(\eta') \frac{\nabla^2 \phi_{s}(\eta',\vec{x})}{4\pi G} \right] \right\rangle \,.
\end{equation}
Integrating by parts inside the spatial average once again leads to,
\begin{equation}
	\label{eq:cosmic_energy_step3}
	\int \frac{d^3 \vec{q}}{(2\pi)^3} q^2 \bar{f}_{l,\textrm{ctr}}^{(2)}(\eta,q) = \int_{0}^{\eta} d\eta' a(\eta') \frac{d}{d\eta'} \left\{a^{2}(\eta') \left\langle \left[\frac{\vec{\nabla} \phi_{s} (\eta',\vec{x})}{4\pi G}\right]^2 \right\rangle \right\} \,.
\end{equation}
Now note that, using Eqs.~(\ref{eq:poisson}) and (\ref{eq:pot}),
\begin{equation}
	\label{eq:cosmic_energy_step4}
	\left\langle \left[\frac{\vec{\nabla} \phi_{s} (\eta',\vec{x})}{4\pi G}\right]^2  \right\rangle  = - \left\langle \frac{\phi_{s}(\eta',\vec{x}) \nabla^2 \phi_{s}(\eta',\vec{x})}{4\pi G} \right\rangle  = - 2a^2(\eta') \bar{\rho}(\eta') u(\eta') \,,
\end{equation}
which yields
\begin{equation}
	\label{eq:cosmic_energy_step5}
	\int \frac{d^3 \vec{q}}{(2\pi)^3} q^2 \bar{f}_{l,\textrm{ctr}}^{(2)}(\eta,q) = -2a^{3}(\eta) \bar{\rho}(\eta) \int_{0}^{\eta} d\eta' a(\eta') \frac{d}{d\eta'} \left[a(\eta') u(\eta')\right] \,,
\end{equation}
where we used the relation $a^{3}(\eta') \bar{\rho}(\eta') = a^{3}(\eta) \bar{\rho}(\eta)$ to bring this quantity out of the integral. This may now be combined with the definition for the short scale kinetic energy per unit mass, Eq.~(\ref{eq:resvelocityrepeat}), to produce the desired result
\begin{equation}
	\label{eq:cosmic_energy_step6}
	\kappa(\eta) = -a^{-2}(\eta) \int_{0}^{\eta} d\eta' a(\eta') \frac{d}{d\eta'} \left[a(\eta') u(\eta') \right] \,.
\end{equation}

\section{One-loop power spectrum}
\label{sec:app4}

In Appendix \ref{sec:app1} we include explicit formulas for the full time-dependent coefficients entering SPT kernels in $\Lambda$CDM. These can be combined with the scale dependencies shown in Eqs.~(\ref{eq:separable-2nd}) and (\ref{eq:scale}) at second order, and Eqs.~(\ref{eq:separable-3rd}) and (\ref{eq:scale-3rd}) at third order, to compute perturbative predictions for cosmological observables of interest. Here we provide explicit formulas, for the one-loop power spectrum, for the reader's convenience. 

The (equal-time) power spectrum, $P(a,k)$ is defined by:
\begin{equation}
\label{eq:power_spectrum}
	\langle \delta(a,\vec{k}) \delta(a,\vec{k}') \rangle = (2\pi)^{3} \delta^{(3)}(\vec{k}+\vec{k}') P(a,k) \,.
\end{equation}
From the perturbative expansion Eq.~(\ref{eq:pertsol}), the one-loop power spectrum in SPT reads,
\begin{equation}
\label{eq:1-loop}
	P_{\textrm{1-loop}}(a,k) = D_{\textrm{L}}^{2}(a)P_{\textrm{L}}(k) + P_{13}(a,k) + P_{22}(a,k) \,.
\end{equation}
We recall that $P_{\textrm{L}}(k)$ is the linear power spectrum today and $D_{\textrm{L}}(a=1)=1$ sets the linear growth factor normalization [see Eq.~(\ref{eq:analytic})]. The other two contributions will be flashed out in what follows. We first have, from Eqs.~(\ref{eq:separable}) and (\ref{eq:separable-3rd}),
\begin{equation}
\label{eq:p13}
	P_{13}(a,k) = 2D_{\textrm{L}}(a) \sum_{i=1}^{6} c_{i}^{(3)}(a) \langle \delta_{\textrm{L}}(\vec{k}) h_{i}^{(3)}(-\vec{k})\rangle' \equiv 2D_{\textrm{L}}(a) \sum_{i=1}^{6} c_{i}^{(3)}(a) \Gamma_{i}(k)  \,,
\end{equation}
where $\Gamma_{i}(k) = \langle \delta_{\textrm{L}}(\vec{k}) h_{i}^{(3)}(-\vec{k})\rangle'$ and the primed correlation function has the momentum conserving Dirac delta stripped off, following standard convention. Formulas for the time-dependent coefficients $c_{i}^{(3)}(a)$ can be found in Appendix \ref{sec:app1}, and from Eq.~(\ref{eq:scale-3rd}) we obtain:
\begin{equation}
\label{eq:p13_scale}
\begin{split}
	& \Gamma_{1}(k) = \Gamma_{2}(k) = -\frac{1}{3} k^3 P_{\textrm{L}}(k) \int_{0}^{\infty} \frac{dx}{2\pi^2} P_{\textrm{L}}(kx) (1+x^{2}) \\ & \Gamma_{3}(k) = -\frac{1}{8} k^3 P_{\textrm{L}}(k) \int_{0}^{\infty} \frac{dx}{2\pi^2} P_{\textrm{L}}(kx) \left[2(x^4-4x^2-1)+\frac{(x^2-1)^{3}}{x} \log \left|\frac{x-1}{x+1}\right| \right] \\ & \Gamma_{4}(k) = \frac{1}{3} k^3 P_{\textrm{L}}(k) \int_{0}^{\infty} \frac{dx}{2\pi^2} P_{\textrm{L}}(kx) x^{2} \\ & \Gamma_{5}(k) = -\frac{1}{16} k^3 P_{\textrm{L}}(k) \int_{0}^{\infty} \frac{dx}{2\pi^2} P_{\textrm{L}}(kx) \frac{1}{x^3} \left[2x(x^4+4x^2-1)+(x^2-1)^{3} \log \left|\frac{x-1}{x+1}\right| \right] \\ & \Gamma_{6}(k) = - \frac{1}{6} k^3 P_{\textrm{L}}(k) \int_{0}^{\infty} \frac{dx}{2\pi^2} P_{\textrm{L}}(kx) \,.
\end{split}
\end{equation}

In the EdS approximation the time-dependent coefficients simplify to Eq.~(\ref{eq:eds_3rd}), and from Eq.~(\ref{eq:p13}) we arrive at
\begin{equation}
\label{eq:p13_eds}
\begin{split}
	P_{13}(a,k) \approx & \frac{1}{84\pi^2} D_{\textrm{L}}^{4}(a) k^{3} P_{\textrm{L}}(k) \int_{0}^{\infty} dx \frac{P_{\textrm{L}}(kx)}{x^{2}}\Bigg[1-\frac{79}{6}x^2 + \frac{25}{3} x^4 \\ & -\frac{7}{2} x^6 - \frac{1}{2x} (x^2-1)^3 (1+\frac{7}{2}x^2) \log \left|\frac{x-1}{x+1}\right| \Bigg] \,.
\end{split}
\end{equation} 
We similarly need, from Eq.(\ref{eq:separable-2nd}),
\begin{equation}
	\label{eq:p22}
	P_{22}(a,k) = \sum_{i=1}^{2} \sum_{j=1}^{2} c_{i}^{(2)}(a) c_{j}^{(2)}(a) \langle h_{i}^{(2)}(\vec{k}) h_{j}^{(2)}(-\vec{k})\rangle' \equiv \sum_{i=1}^{2} \sum_{j=1}^{2} c_{i}^{(2)}(a) c_{j}^{(2)}(a) \Sigma_{ij}(k) \,,
\end{equation}
where $\Sigma_{ij}(k) =  \langle h_{i}^{(2)}(\vec{k}) h_{j}^{(2)}(-\vec{k})\rangle'$ [see Appendix \ref{sec:app1} for the time-dependent coefficients $c_{i}^{(2)}(a)$]. We obtain from Eq.~(\ref{eq:scale})
\begin{equation}
	\label{eq:p22_scale}
	\begin{split}
		&  \Sigma_{11}(k) = \frac{1}{2} k^3 \int_{0}^{\infty} \frac{dx}{2\pi^2} \int_{-1}^{1} \frac{dt}{2} \frac{(t+x-2xt^2)^2}{(1-2xt+x^2)^2} P_{\textrm{L}}(kx) P_{\textrm{L}}\left(k\sqrt{1-2xt+x^2}\right) \\ & \Sigma_{12}(k) =  \Sigma_{21}(k) = \frac{1}{2} k^{3} \int_{0}^{\infty} \frac{dx}{2\pi^2} \int_{-1}^{1} \frac{dt}{2} \frac{(t+x-2xt^2)(t-x)}{(1-2xt+x^2)^2} P_{\textrm{L}}(kx) P_{\textrm{L}}\left(k\sqrt{1-2xt+x^2}\right) \\ & \Sigma_{22}(k) = \frac{1}{2} k^3 \int_{0}^{\infty} \frac{dx}{2\pi^2} \int_{-1}^{1} \frac{dt}{2} \frac{(t-x)^2}{(1-2xt+x^2)^2} P_{\textrm{L}}(kx) P_{\textrm{L}}\left(k\sqrt{1-2xt+x^2}\right) \,.
	\end{split}
\end{equation}

In the EdS approximation the time-dependent coefficients simplify to Eq.~(\ref{eq:eds_2nd}), and from Eq.~(\ref{eq:p22}) we arrive at
\begin{equation}
	\label{eq:p22_eds}
	\begin{split}
		P_{22}(a,k) \approx & \frac{1}{392\pi^2} D_{\textrm{L}}^{4}(a) k^{3} \int_{0}^{\infty} dx \int_{-1}^{1} dt \frac{(7t+3x-10t^2x)^2}{(1-2xt+x^2)^2} P_{\textrm{L}}(kx) P_{\textrm{L}}\left(k\sqrt{1-2xt+x^2}\right)  \,.
	\end{split}
\end{equation} 
\section{Tree-level bispectrum}
\label{sec:app5}

In this section we explicitly write down the relevant formulas for the tree-level bispectrum in SPT, with the full time dependencies of $\Lambda$CDM kernels. The bispectrum $B(k_1,k_2,k_3)$ is defined by
\begin{equation}
\label{eq:bispectrum}
	\langle \delta(\vec{k_1}) \delta(\vec{k_2}) \delta(\vec{k_3}) \rangle = (2\pi)^{3}\delta^{(3)}(\vec{k_1}+\vec{k_2}+\vec{k_3}) B(k_1,k_2,k_3) \,.
\end{equation}
From Eqs.~(\ref{eq:pertsol}), (\ref{eq:separable}), and (\ref{eq:separable-2nd}) the tree-level bispectrum in SPT reads
\begin{equation}
\label{eq:tree_bispectrum}
	B(k_1,k_2,k_3) = D_{\textrm{L}}^{2}(a) \sum_{i=1}^{2} c_{i}^{2}(a) \langle h_{i}^{(2)}(\vec{k}_1) \delta_{\textrm{L}}(\vec{k}_2) \delta_{\textrm{L}}(\vec{k}_3) \rangle' + \dots \,,
\end{equation}
where the ellipsis represent two additional terms corresponding to permutations of the three wavenumbers. The scale dependent part $\langle h_{i}^{(2)}(\vec{k}_1) \delta_{\textrm{L}}(\vec{k}_2) \delta_{\textrm{L}}(\vec{k}_3) \rangle'$ can be computed from Eq.~(\ref{eq:scale}), and Eq.~(\ref{eq:tree_bispectrum}) becomes
\begin{equation}
\label{eq:tree_bispectrum_final}
	B(k_1,k_2,k_3) = 2D_{\textrm{L}}^{2}(a) \left[c_{1}^{(2)}(a)\alpha^{(s)}(\vec{k}_{1}, \vec{k}_{2}) + c_{2}^{(2)}(a)\beta(\vec{k}_{1}, \vec{k}_{2}) \right]P_{\textrm{L}}(k_1) P_{\textrm{L}}(k_2) + \dots \,,
\end{equation}
where
\begin{equation}
\label{eq:alpha_beta}
\begin{split}
	& \alpha^{(s)}(\vec{k}_{1}, \vec{k}_{2}) = \frac{1}{2} \frac{\vec{k}_{1} \cdot \vec{k}_{12}}{k_{1}^2} + \frac{1}{2} \frac{\vec{k}_{2} \cdot \vec{k}_{12}}{k_{2}^2} \\ & \beta(\vec{k}_{1}, \vec{k}_{2}) = \frac{k_{12}^2 (\vec{k}_1 \cdot \vec{k}_2)}{2k_1^2 k_2^2} \,,
\end{split}
\end{equation}
with $\vec{k}_{12} = \vec{k}_1+\vec{k}_2$, and formulas for the time-dependent coefficients $c_{i}^{(2)}(a)$ can be found in Appendix \ref{sec:app1}. In the EdS approximation they simplify to Eq.~(\ref{eq:eds_2nd}), and from Eq.(\ref{eq:tree_bispectrum_final}) we arrive at
\begin{equation}
\label{eq:tree_bispectrum_eds}
	B(k_1,k_2,k_3) \approx 2D_{\textrm{L}}^{4}(a) F_{2}(\vec{k}_1,\vec{k}_2) P_{\textrm{L}}(k_1) P_{\textrm{L}}(k_2) + \dots \,,
\end{equation}
where
\begin{equation}
\label{eq:eds_kernel}
\begin{split}
	F_{2}(\vec{k}_1,\vec{k}_2) &= \frac{5}{7} \alpha^{(s)}(\vec{k}_{1}, \vec{k}_{2}) + \frac{2}{7} \beta(\vec{k}_{1}, \vec{k}_{2}) \\ & = \frac{5}{7} + \frac{1}{2} \frac{\vec{k}_1 \cdot \vec{k}_2}{k_1k_2} \left(\frac{k_1}{k_2}+ \frac{k_2}{k_1}\right) + \frac{2}{7} \frac{(\vec{k}_1 \cdot \vec{k}_2)^2}{k_1^2k_2^2} \,.
\end{split}
\end{equation}

\bibliography{cpt_ps.bib}

\begin{thebibliography}{100}

\bibitem{Makino:1991rp}
Nobuyoshi Makino, Misao Sasaki, and Yasushi Suto.
\newblock {Analytic approach to the perturbative expansion of nonlinear
  gravitational fluctuations in logical density and velocity fields}.
\newblock {\em Phys. Rev. D}, 46:585--602, 1992.

\bibitem{Jain:1993jh}
Bhuvnesh Jain and Edmund Bertschinger.
\newblock {Second order power spectrum and nonlinear evolution at high
  redshift}.
\newblock {\em Astrophys. J.}, 431:495, 1994.

\bibitem{Goroff:1986ep}
M.~H. Goroff, Benjamin Grinstein, S.~J. Rey, and Mark~B. Wise.
\newblock {Coupling of Modes of Cosmological Mass Density Fluctuations}.
\newblock {\em Astrophys. J.}, 311:6--14, 1986.

\bibitem{Zeldovich:1969sb}
Ya.~B. Zeldovich.
\newblock {Gravitational instability: An Approximate theory for large density
  perturbations}.
\newblock {\em Astron. Astrophys.}, 5:84--89, 1970.

\bibitem{Bernardeau:2001qr}
F.~Bernardeau, S.~Colombi, E.~Gaztanaga, and R.~Scoccimarro.
\newblock {Large scale structure of the universe and cosmological perturbation
  theory}.
\newblock {\em Phys. Rept.}, 367:1--248, 2002.

\bibitem{Scoccimarro:1995if}
Roman Scoccimarro and Joshua Frieman.
\newblock {Loop corrections in nonlinear cosmological perturbation theory}.
\newblock {\em Astrophys. J. Suppl.}, 105:37, 1996.

\bibitem{Bottaro:2023wkd}
Salvatore Bottaro, Emanuele Castorina, Marco Costa, Diego Redigolo, and Ennio
  Salvioni.
\newblock {Unveiling Dark Forces with Measurements of the Large Scale Structure
  of the Universe}.
\newblock {\em Phys. Rev. Lett.}, 132(20):201002, 2024.

\bibitem{Piga:2022mge}
Lorenzo Piga, Marco Marinucci, Guido D'Amico, Massimo Pietroni, Filippo
  Vernizzi, and Bill~S. Wright.
\newblock {Constraints on modified gravity from the BOSS galaxy survey}.
\newblock {\em JCAP}, 04:038, 2023.

\bibitem{Lewandowski:2019txi}
Matthew Lewandowski.
\newblock {Violation of the consistency relations for large-scale structure
  with dark energy}.
\newblock {\em JCAP}, 08:044, 2020.

\bibitem{Crisostomi:2019vhj}
Marco Crisostomi, Matthew Lewandowski, and Filippo Vernizzi.
\newblock {Consistency relations for large-scale structure in modified gravity
  and the matter bispectrum}.
\newblock {\em Phys. Rev. D}, 101(12):123501, 2020.

\bibitem{Bottaro:2024pcb}
Salvatore Bottaro, Emanuele Castorina, Marco Costa, Diego Redigolo, and Ennio
  Salvioni.
\newblock {From 100 kpc to 10 Gpc: Dark Matter self-interactions before and
  after DESI}.
\newblock 7 2024.

\bibitem{Schneider:2015yka}
Aurel Schneider, Romain Teyssier, Doug Potter, Joachim Stadel, Julian Onions,
  Darren~S. Reed, Robert~E. Smith, Volker Springel, Frazer~R. Pearce, and Roman
  Scoccimarro.
\newblock {Matter power spectrum and the challenge of percent accuracy}.
\newblock {\em JCAP}, 04:047, 2016.

\bibitem{Carlson:2009it}
Jordan Carlson, Martin White, and Nikhil Padmanabhan.
\newblock {A critical look at cosmological perturbation theory techniques}.
\newblock {\em Phys. Rev. D}, 80:043531, 2009.

\bibitem{Buchert:1993ud}
Thomas Buchert.
\newblock {Lagrangian theory of gravitational instability of Friedman-Lemaitre
  cosmologies: Generic third order model for nonlinear clustering}.
\newblock {\em Mon. Not. Roy. Astron. Soc.}, 267:811--820, 1994.

\bibitem{Catelan:1994ze}
Paolo Catelan.
\newblock {Lagrangian dynamics in nonflat universes and nonlinear gravitational
  evolution}.
\newblock {\em Mon. Not. Roy. Astron. Soc.}, 276:115, 1995.

\bibitem{Bouchet:1994xp}
F.~R. Bouchet, S.~Colombi, E.~Hivon, and R.~Juszkiewicz.
\newblock {Perturbative Lagrangian approach to gravitational instability}.
\newblock {\em Astron. Astrophys.}, 296:575, 1995.

\bibitem{Sugiyama:2013mpa}
Naonori~S. Sugiyama.
\newblock {Using Lagrangian perturbation theory for precision cosmology}.
\newblock {\em Astrophys. J.}, 788:63, 2014.

\bibitem{Vlah:2015zda}
Zvonimir Vlah, Uro\v{s} Seljak, Man~Yat Chu, and Yu~Feng.
\newblock {Perturbation theory, effective field theory, and oscillations in the
  power spectrum}.
\newblock {\em JCAP}, 03:057, 2016.

\bibitem{Euclid:2023tog}
A.~Pezzotta et~al.
\newblock {Euclid preparation. TBD. Galaxy power spectrum modelling in real
  space}.
\newblock 12 2023.

\bibitem{Chudaykin:2020aoj}
Anton Chudaykin, Mikhail~M. Ivanov, Oliver H.~E. Philcox, and Marko
  Simonovi\'c.
\newblock {Nonlinear perturbation theory extension of the Boltzmann code
  CLASS}.
\newblock {\em Phys. Rev. D}, 102(6):063533, 2020.

\bibitem{Ivanov:2019pdj}
Mikhail~M. Ivanov, Marko Simonovi\'c, and Matias Zaldarriaga.
\newblock {Cosmological Parameters from the BOSS Galaxy Power Spectrum}.
\newblock {\em JCAP}, 05:042, 2020.

\bibitem{Philcox:2022frc}
Oliver H.~E. Philcox, Mikhail~M. Ivanov, Giovanni Cabass, Marko Simonovi\'c,
  Matias Zaldarriaga, and Takahiro Nishimichi.
\newblock {Cosmology with the redshift-space galaxy bispectrum monopole at
  one-loop order}.
\newblock {\em Phys. Rev. D}, 106(4):043530, 2022.

\bibitem{Taule:2023izt}
Petter Taule and Mathias Garny.
\newblock {The two-loop power spectrum in redshift space}.
\newblock {\em JCAP}, 11:078, 2023.

\bibitem{Beutler:2011hx}
Florian Beutler, Chris Blake, Matthew Colless, D.~Heath Jones, Lister
  Staveley-Smith, Lachlan Campbell, Quentin Parker, Will Saunders, and Fred
  Watson.
\newblock {The 6dF Galaxy Survey: Baryon Acoustic Oscillations and the Local
  Hubble Constant}.
\newblock {\em Mon. Not. Roy. Astron. Soc.}, 416:3017--3032, 2011.

\bibitem{Blake:2011rj}
Chris Blake et~al.
\newblock {The WiggleZ Dark Energy Survey: the growth rate of cosmic structure
  since redshift z=0.9}.
\newblock {\em Mon. Not. Roy. Astron. Soc.}, 415:2876, 2011.

\bibitem{Chudaykin:2022nru}
Anton Chudaykin and Mikhail~M. Ivanov.
\newblock {Cosmological constraints from the power spectrum of eBOSS quasars}.
\newblock {\em Phys. Rev. D}, 107(4):043518, 2023.

\bibitem{Taule:2024bot}
Petter Taule, Marco Marinucci, Giorgia Biselli, Massimo Pietroni, and Filippo
  Vernizzi.
\newblock {Constraints on dark energy and modified gravity from the BOSS
  Full-Shape and DESI BAO data}.
\newblock 9 2024.

\bibitem{Aviles:2024zlw}
Alejandro Aviles.
\newblock {Testing gravity with the full-shape galaxy power spectrum: first
  constraints on scale-dependent modified gravity}.
\newblock 9 2024.

\bibitem{Colas:2019ret}
Thomas Colas, Guido D'amico, Leonardo Senatore, Pierre Zhang, and Florian
  Beutler.
\newblock {Efficient Cosmological Analysis of the SDSS/BOSS data from the
  Effective Field Theory of Large-Scale Structure}.
\newblock {\em JCAP}, 06:001, 2020.

\bibitem{Simon:2022csv}
Th\'eo Simon, Pierre Zhang, and Vivian Poulin.
\newblock {Cosmological inference from the EFTofLSS: the eBOSS QSO full-shape
  analysis}.
\newblock {\em JCAP}, 07:041, 2023.

\bibitem{DAmico:2022osl}
Guido D'Amico, Yaniv Donath, Matthew Lewandowski, Leonardo Senatore, and Pierre
  Zhang.
\newblock {The BOSS bispectrum analysis at one loop from the Effective Field
  Theory of Large-Scale Structure}.
\newblock {\em JCAP}, 05:059, 2024.

\bibitem{DESI:2024jis}
A.~G. Adame et~al.
\newblock {DESI 2024 V: Full-Shape Galaxy Clustering from Galaxies and
  Quasars}.
\newblock 11 2024.

\bibitem{Chen:2024vvk}
S.~Chen et~al.
\newblock {Analysis of DESI\texttimes{}DES using the Lagrangian effective
  theory of LSS}.
\newblock {\em Phys. Rev. D}, 110(10):103518, 2024.

\bibitem{Chen:2022jzq}
Shi-Fan Chen, Martin White, Joseph DeRose, and Nickolas Kokron.
\newblock {Cosmological analysis of three-dimensional BOSS galaxy clustering
  and Planck CMB lensing cross correlations via Lagrangian perturbation
  theory}.
\newblock {\em JCAP}, 07(07):041, 2022.

\bibitem{Moretti:2023drg}
Chiara Moretti, Maria Tsedrik, Pedro Carrilho, and Alkistis Pourtsidou.
\newblock {Modified gravity and massive neutrinos: constraints from the full
  shape analysis of BOSS galaxies and forecasts for Stage IV surveys}.
\newblock {\em JCAP}, 12:025, 2023.

\bibitem{Crocce:2005xy}
Martin Crocce and Roman Scoccimarro.
\newblock {Renormalized cosmological perturbation theory}.
\newblock {\em Phys. Rev. D}, 73:063519, 2006.

\bibitem{Crocce:2007dt}
Martin Crocce and Roman Scoccimarro.
\newblock {Nonlinear Evolution of Baryon Acoustic Oscillations}.
\newblock {\em Phys. Rev. D}, 77:023533, 2008.

\bibitem{Taruya:2007xy}
Atsushi Taruya and Takashi Hiramatsu.
\newblock {A Closure Theory for Non-linear Evolution of Cosmological Power
  Spectra}.
\newblock {\em Astrophys. J.}, 674:617, 2008.

\bibitem{Bernardeau:2008fa}
Francis Bernardeau, Martin Crocce, and Roman Scoccimarro.
\newblock {Multi-Point Propagators in Cosmological Gravitational Instability}.
\newblock {\em Phys. Rev. D}, 78:103521, 2008.

\bibitem{Taruya:2012ut}
Atsushi Taruya, Francis Bernardeau, Takahiro Nishimichi, and Sandrine Codis.
\newblock {RegPT: Direct and fast calculation of regularized cosmological power
  spectrum at two-loop order}.
\newblock {\em Phys. Rev. D}, 86:103528, 2012.

\bibitem{Ivanov:2019dlz}
Mikhail~M. Ivanov.
\newblock {\em {Precision theoretical methods for large-scale structure of the
  Universe}}.
\newblock PhD thesis, EPFL, 1 2019.

\bibitem{Sugiyama:2024eye}
Naonori Sugiyama.
\newblock {Developing a Theoretical Model for the Resummation of Infrared
  Effects in the Post-Reconstruction Power Spectrum (youtu.be/u1-xx3\_4xCg)}.
\newblock 2 2024.

\bibitem{Baldauf:2015xfa}
Tobias Baldauf, Mehrdad Mirbabayi, Marko Simonovi\'c, and Matias Zaldarriaga.
\newblock {Equivalence Principle and the Baryon Acoustic Peak}.
\newblock {\em Phys. Rev. D}, 92(4):043514, 2015.

\bibitem{Lewandowski:2018ywf}
Matthew Lewandowski and Leonardo Senatore.
\newblock {An analytic implementation of the IR-resummation for the BAO peak}.
\newblock {\em JCAP}, 03:018, 2020.

\bibitem{Blas:2016sfa}
Diego Blas, Mathias Garny, Mikhail~M. Ivanov, and Sergey Sibiryakov.
\newblock {Time-Sliced Perturbation Theory II: Baryon Acoustic Oscillations and
  Infrared Resummation}.
\newblock {\em JCAP}, 07:028, 2016.

\bibitem{Pietroni:2008jx}
Massimo Pietroni.
\newblock {Flowing with Time: a New Approach to Nonlinear Cosmological
  Perturbations}.
\newblock {\em JCAP}, 10:036, 2008.

\bibitem{Chen:2024pyp}
Shi-Fan Chen, Zvonimir Vlah, and Martin White.
\newblock {The bispectrum in Lagrangian perturbation theory}.
\newblock {\em JCAP}, 11:012, 2024.

\bibitem{Chen:2020zjt}
Shi-Fan Chen, Zvonimir Vlah, Emanuele Castorina, and Martin White.
\newblock {Redshift-Space Distortions in Lagrangian Perturbation Theory}.
\newblock {\em JCAP}, 03:100, 2021.

\bibitem{Matsubara:2007wj}
Takahiko Matsubara.
\newblock {Resumming Cosmological Perturbations via the Lagrangian Picture:
  One-loop Results in Real Space and in Redshift Space}.
\newblock {\em Phys. Rev. D}, 77:063530, 2008.

\bibitem{Baumann:2010tm}
Daniel Baumann, Alberto Nicolis, Leonardo Senatore, and Matias Zaldarriaga.
\newblock {Cosmological Non-Linearities as an Effective Fluid}.
\newblock {\em JCAP}, 07:051, 2012.

\bibitem{Carrasco:2012cv}
John Joseph~M. Carrasco, Mark~P. Hertzberg, and Leonardo Senatore.
\newblock {The Effective Field Theory of Cosmological Large Scale Structures}.
\newblock {\em JHEP}, 09:082, 2012.

\bibitem{Carroll:2013oxa}
Sean~M. Carroll, Stefan Leichenauer, and Jason Pollack.
\newblock {Consistent effective theory of long-wavelength cosmological
  perturbations}.
\newblock {\em Phys. Rev. D}, 90(2):023518, 2014.

\bibitem{Pietroni:2011iz}
Massimo Pietroni, Gianpiero Mangano, Ninetta Saviano, and Matteo Viel.
\newblock {Coarse-Grained Cosmological Perturbation Theory}.
\newblock {\em JCAP}, 01:019, 2012.

\bibitem{Carrasco:2013mua}
John Joseph~M. Carrasco, Simon Foreman, Daniel Green, and Leonardo Senatore.
\newblock {The Effective Field Theory of Large Scale Structures at Two Loops}.
\newblock {\em JCAP}, 07:057, 2014.

\bibitem{Porto:2013qua}
Rafael~A. Porto, Leonardo Senatore, and Matias Zaldarriaga.
\newblock {The Lagrangian-space Effective Field Theory of Large Scale
  Structures}.
\newblock {\em JCAP}, 05:022, 2014.

\bibitem{Vlah:2015sea}
Zvonimir Vlah, Martin White, and Alejandro Aviles.
\newblock {A Lagrangian effective field theory}.
\newblock {\em JCAP}, 09:014, 2015.

\bibitem{Zaldarriaga:2015jrj}
Matias Zaldarriaga and Mehrdad Mirbabayi.
\newblock {Lagrangian Formulation of the Eulerian-EFT}.
\newblock 11 2015.

\bibitem{Baldauf:2016sjb}
Tobias Baldauf, Mehrdad Mirbabayi, Marko Simonovi\'c, and Matias Zaldarriaga.
\newblock {LSS constraints with controlled theoretical uncertainties}.
\newblock 2 2016.

\bibitem{Braganca:2023pcp}
Diogo Braganca, Yaniv Donath, Leonardo Senatore, and Henry Zheng.
\newblock {Peeking into the next decade in Large-Scale Structure Cosmology with
  its Effective Field Theory}.
\newblock 7 2023.

\bibitem{Nishimichi:2020tvu}
Takahiro Nishimichi, Guido D'Amico, Mikhail~M. Ivanov, Leonardo Senatore, Marko
  Simonovi\'c, Masahiro Takada, Matias Zaldarriaga, and Pierre Zhang.
\newblock {Blinded challenge for precision cosmology with large-scale
  structure: results from effective field theory for the redshift-space galaxy
  power spectrum}.
\newblock {\em Phys. Rev. D}, 102(12):123541, 2020.

\bibitem{Konstandin:2019bay}
Thomas Konstandin, Rafael~A. Porto, and Henrique Rubira.
\newblock {The effective field theory of large scale structure at three loops}.
\newblock {\em JCAP}, 11:027, 2019.

\bibitem{Desjacques:2016bnm}
Vincent Desjacques, Donghui Jeong, and Fabian Schmidt.
\newblock {Large-Scale Galaxy Bias}.
\newblock {\em Phys. Rept.}, 733:1--193, 2018.

\bibitem{Baldauf:2014qfa}
Tobias Baldauf, Lorenzo Mercolli, Mehrdad Mirbabayi, and Enrico Pajer.
\newblock {The Bispectrum in the Effective Field Theory of Large Scale
  Structure}.
\newblock {\em JCAP}, 05:007, 2015.

\bibitem{Baldauf:2021zlt}
Tobias Baldauf, Mathias Garny, Petter Taule, and Theo Steele.
\newblock {Two-loop bispectrum of large-scale structure}.
\newblock {\em Phys. Rev. D}, 104(12):123551, 2021.

\bibitem{Bertolini:2016bmt}
Daniele Bertolini, Katelin Schutz, Mikhail~P. Solon, and Kathryn~M. Zurek.
\newblock {The Trispectrum in the Effective Field Theory of Large Scale
  Structure}.
\newblock {\em JCAP}, 06:052, 2016.

\bibitem{Euclid:2020rfv}
M.~Knabenhans et~al.
\newblock {Euclid preparation: IX. EuclidEmulator2 \textendash{} power spectrum
  emulation with massive neutrinos and self-consistent dark energy
  perturbations}.
\newblock {\em Mon. Not. Roy. Astron. Soc.}, 505(2):2840--2869, 2021.

\bibitem{Angulo:2020vky}
Raul~E. Angulo, Matteo Zennaro, Sergio Contreras, Giovanni Aric\`o, Marcos
  Pellejero-Iba\~nez, and Jens St\"ucker.
\newblock {The BACCO simulation project: exploiting the full power of
  large-scale structure for cosmology}.
\newblock {\em Mon. Not. Roy. Astron. Soc.}, 507(4):5869--5881, 2021.

\bibitem{Heitmann:2013bra}
Katrin Heitmann, Earl Lawrence, Juliana Kwan, Salman Habib, and David Higdon.
\newblock {The Coyote Universe Extended: Precision Emulation of the Matter
  Power Spectrum}.
\newblock {\em Astrophys. J.}, 780:111, 2014.

\bibitem{Nishimichi:2018etk}
Takahiro Nishimichi et~al.
\newblock {Dark Quest. I. Fast and Accurate Emulation of Halo Clustering
  Statistics and Its Application to Galaxy Clustering}.
\newblock {\em Astrophys. J.}, 884:29, 2019.

\bibitem{DeRose:2018xdj}
Joseph DeRose, Risa~H. Wechsler, Jeremy~L. Tinker, Matthew~R. Becker, Yao-Yuan
  Mao, Thomas McClintock, Sean McLaughlin, Eduardo Rozo, and Zhongxu Zhai.
\newblock {The Aemulus Project I: Numerical Simulations for Precision
  Cosmology}.
\newblock {\em Astrophys. J.}, 875(1):69, 2019.

\bibitem{Wibking:2017slg}
Benjamin~D. Wibking, Andr\'es~N. Salcedo, David~H. Weinberg, Lehman~H.
  Garrison, Douglas Ferrer, Jeremy Tinker, Daniel Eisenstein, Marc Metchnik,
  and Philip Pinto.
\newblock {Emulating galaxy clustering and galaxy\textendash{}galaxy lensing
  into the deeply non-linear regime: methodology, information, and forecasts}.
\newblock {\em Mon. Not. Roy. Astron. Soc.}, 484(1):989--1006, 2019.

\bibitem{Winther:2019mus}
Hans Winther, Santiago Casas, Marco Baldi, Kazuya Koyama, Baojiu Li, Lucas
  Lombriser, and Gong-Bo Zhao.
\newblock {Emulators for the nonlinear matter power spectrum beyond
  $\Lambda$CDM}.
\newblock {\em Phys. Rev. D}, 100(12):123540, 2019.

\bibitem{Brando:2022gvg}
Guilherme Brando, Bartolomeo Fiorini, Kazuya Koyama, and Hans~A. Winther.
\newblock {Enabling matter power spectrum emulation in
  beyond-\ensuremath{\Lambda}CDM cosmologies with COLA}.
\newblock {\em JCAP}, 09:051, 2022.

\bibitem{Mauland:2023pjt}
Renate Mauland, Hans~A. Winther, and Cheng-Zong Ruan.
\newblock {Sesame: A power spectrum emulator pipeline for
  beyond-\ensuremath{\Lambda}CDM models}.
\newblock {\em Astron. Astrophys.}, 685:A156, 2024.

\bibitem{Jamieson:2024fsp}
Drew Jamieson, Yin Li, Francisco Villaescusa-Navarro, Shirley Ho, and David~N.
  Spergel.
\newblock {Field-level Emulation of Cosmic Structure Formation with Cosmology
  and Redshift Dependence}.
\newblock 8 2024.

\bibitem{Jamieson:2022lqc}
Drew Jamieson, Yin Li, Renan~Alves de~Oliveira, Francisco Villaescusa-Navarro,
  Shirley Ho, and David~N. Spergel.
\newblock {Field-level Neural Network Emulator for Cosmological N-body
  Simulations}.
\newblock {\em Astrophys. J.}, 952(2):145, 2023.

\bibitem{Garny:2022tlk}
Mathias Garny, Dominik Laxhuber, and Roman Scoccimarro.
\newblock {Perturbation theory with dispersion and higher cumulants: Framework
  and linear theory}.
\newblock {\em Phys. Rev. D}, 107(6):063539, 2023.

\bibitem{Garny:2022kbk}
Mathias Garny, Dominik Laxhuber, and Roman Scoccimarro.
\newblock {Perturbation theory with dispersion and higher cumulants: Nonlinear
  regime}.
\newblock {\em Phys. Rev. D}, 107(6):063540, 2023.

\bibitem{Garny:2025ovs}
Mathias Garny and Roman Scoccimarro.
\newblock {Vlasov Perturbation Theory and the role of higher cumulants}.
\newblock 2 2025.

\bibitem{Erschfeld:2018zqg}
Alaric Erschfeld and Stefan Floerchinger.
\newblock {Evolution of dark matter velocity dispersion}.
\newblock {\em JCAP}, 06:039, 2019.

\bibitem{Dominguez:2000dt}
Alvaro Dominguez.
\newblock {Hydrodynamic approach to the evolution of cosmological structures}.
\newblock 9 2000.

\bibitem{mcdonald2011generate}
Patrick McDonald.
\newblock How to generate a significant effective temperature for cold dark
  matter, from first principles.
\newblock {\em Journal of Cosmology and Astroparticle Physics}, 2011(04):032,
  2011.

\bibitem{Kitaura:2023mfc}
F.~S. Kitaura, F.~Sinigaglia, A.~Balaguera-Antol\'\i{}nez, and G.~Favole.
\newblock {The Cosmic Web from Perturbation Theory}.
\newblock 1 2023.

\bibitem{Matarrese:2007wc}
Sabino Matarrese and Massimo Pietroni.
\newblock {Resumming Cosmic Perturbations}.
\newblock {\em JCAP}, 06:026, 2007.

\bibitem{McDonald:2017ths}
Patrick McDonald and Zvonimir Vlah.
\newblock {Large-scale structure perturbation theory without losing stream
  crossing}.
\newblock {\em Phys. Rev. D}, 97(2):023508, 2018.

\bibitem{Aviles:2015osc}
Alejandro Aviles.
\newblock {Dark matter dispersion tensor in perturbation theory}.
\newblock {\em Phys. Rev. D}, 93:063517, 2016.

\bibitem{Bharadwaj:1995px}
Somnath Bharadwaj.
\newblock {Perturbative growth of cosmological clustering. 2: The Two point
  correlation}.
\newblock {\em Astrophys. J.}, 460:28--50, 1996.

\bibitem{Valageas:2001df}
P.~Valageas.
\newblock {Dynamics of gravitational clustering I. building perturbative
  expansions}.
\newblock {\em Astron. Astrophys.}, 379:8, 2001.

\bibitem{tassev2011helmholtz}
Svetlin~V Tassev.
\newblock The helmholtz hierarchy: phase space statistics of cold dark matter.
\newblock {\em Journal of Cosmology and Astroparticle Physics}, 2011(10):022,
  2011.

\bibitem{Valageas:2003gm}
P.~Valageas.
\newblock {A new approach to gravitational clustering: a path-integral
  formalism and large-n expansions}.
\newblock {\em Astron. Astrophys.}, 421:23--40, 2004.

\bibitem{Pueblas:2008uv}
Sebastian Pueblas and Roman Scoccimarro.
\newblock {Generation of Vorticity and Velocity Dispersion by Orbit Crossing}.
\newblock {\em Phys. Rev. D}, 80:043504, 2009.

\bibitem{doi:10.1080/03091928208209001}
S.~F.~Shandarin V.~I.~Arnold and Ya.~B. Zeldovich.
\newblock The large scale structure of the universe i. general properties.
  one-and two-dimensional models.
\newblock {\em Geophysical \& Astrophysical Fluid Dynamics}, 20(1-2):111--130,
  1982.

\bibitem{Musso:2024roa}
Marcello Musso, Giulia Despali, and Ravi~K. Sheth.
\newblock {The energy shear of protohaloes}.
\newblock 5 2024.

\bibitem{Jaber:2023rjx}
Mariana Jaber, Marius Peper, Wojciech~A. Hellwing, Miguel~Angel Aragon-Calvo,
  and Octavio Valenzuela.
\newblock {Hierarchical structure of the cosmic web and galaxy properties}.
\newblock 4 2023.

\bibitem{Aragon-Calvo:2023oxm}
M.~A. Aragon-Calvo.
\newblock {Hierarchical Reconstruction of the Cosmic Web, The H-Spine method}.
\newblock 8 2023.

\bibitem{Libeskind:2017tun}
Noam~I. Libeskind et~al.
\newblock {Tracing the cosmic web}.
\newblock {\em Mon. Not. Roy. Astron. Soc.}, 473(1):1195--1217, 2018.

\bibitem{Cautun:2014fwa}
Marius Cautun, Rien van~de Weygaert, Bernard J.~T. Jones, and Carlos~S. Frenk.
\newblock {Evolution of the cosmic web}.
\newblock {\em Mon. Not. Roy. Astron. Soc.}, 441(4):2923--2973, 2014.

\bibitem{Kugel:2024zxq}
Roi Kugel and Rien van~de Weygaert.
\newblock {Cosmic Web Dynamics: Forces and Strains}.
\newblock 7 2024.

\bibitem{Buehlmann:2018qmm}
Michael Buehlmann and Oliver Hahn.
\newblock {Large-Scale Velocity Dispersion and the Cosmic Web}.
\newblock {\em Mon. Not. Roy. Astron. Soc.}, 487(1):228--245, 2019.

\bibitem{Jelic-Cizmek:2018gdp}
Goran Jelic-Cizmek, Francesca Lepori, Julian Adamek, and Ruth Durrer.
\newblock {The generation of vorticity in cosmological N-body simulations}.
\newblock {\em JCAP}, 09:006, 2018.

\bibitem{Pichon:1999tk}
C.~Pichon and F.~Bernardeau.
\newblock {Vorticity generation in large scale structure caustics}.
\newblock {\em Astron. Astrophys.}, 343:663, 1999.

\bibitem{Senatore:2017hyk}
Leonardo Senatore and Matias Zaldarriaga.
\newblock {The Effective Field Theory of Large-Scale Structure in the presence
  of Massive Neutrinos}.
\newblock 7 2017.

\bibitem{Pajer:2017ulp}
Enrico Pajer and Drian van~der Woude.
\newblock {Divergence of Perturbation Theory in Large Scale Structures}.
\newblock {\em JCAP}, 05:039, 2018.

\bibitem{Blas:2011rf}
Diego Blas, Julien Lesgourgues, and Thomas Tram.
\newblock {The Cosmic Linear Anisotropy Solving System (CLASS) II:
  Approximation schemes}.
\newblock {\em JCAP}, 07:034, 2011.

\bibitem{Hamilton:2000tk}
A.~J.~S. Hamilton.
\newblock {Formulae for growth factors in expanding universes containing matter
  and a cosmological constant}.
\newblock {\em Mon. Not. Roy. Astron. Soc.}, 322:419, 2001.

\bibitem{Hartmeier:2023brx}
Michael Hartmeier and Mathias Garny.
\newblock {Minimal basis for exact time dependent kernels in cosmological
  perturbation theory and application to \ensuremath{\Lambda}CDM and w $_{0}$
  waCDM}.
\newblock {\em JCAP}, 12:027, 2023.

\bibitem{Takahashi:2008yk}
Ryuichi Takahashi.
\newblock {Third Order Density Perturbation and One-loop Power Spectrum in a
  Dark Energy Dominated Universe}.
\newblock {\em Prog. Theor. Phys.}, 120:549--559, 2008.

\bibitem{Bernardeau:1993qu}
Francis Bernardeau.
\newblock {Skewness and Kurtosis in large scale cosmic fields}.
\newblock {\em Astrophys. J.}, 433:1, 1994.

\bibitem{Fasiello:2022lff}
Matteo Fasiello, Tomohiro Fujita, and Zvonimir Vlah.
\newblock {Perturbation theory of large scale structure in the
  \ensuremath{\Lambda}CDM Universe: Exact time evolution and the two-loop power
  spectrum}.
\newblock {\em Phys. Rev. D}, 106(12):123504, 2022.

\bibitem{Fasiello:2016qpn}
Matteo Fasiello and Zvonimir Vlah.
\newblock {Nonlinear fields in generalized cosmologies}.
\newblock {\em Phys. Rev. D}, 94(6):063516, 2016.

\bibitem{Choustikov:2023uyk}
Nicholas Choustikov, Zvonimir Vlah, and Anthony Challinor.
\newblock {Optimizing the evolution of perturbations in the
  \ensuremath{\Lambda}CDM universe}.
\newblock {\em Phys. Rev. D}, 108(2):023529, 2023.

\bibitem{Lewandowski:2016yce}
Matthew Lewandowski, Azadeh Maleknejad, and Leonardo Senatore.
\newblock {An effective description of dark matter and dark energy in the
  mildly non-linear regime}.
\newblock {\em JCAP}, 05:038, 2017.

\bibitem{Fujita:2020xtd}
Tomohiro Fujita and Zvonimir Vlah.
\newblock {Perturbative description of biased tracers using consistency
  relations of LSS}.
\newblock {\em JCAP}, 10:059, 2020.

\bibitem{Donath:2020abv}
Yaniv Donath and Leonardo Senatore.
\newblock {Biased Tracers in Redshift Space in the EFTofLSS with exact time
  dependence}.
\newblock {\em JCAP}, 10:039, 2020.

\bibitem{Schmidt:2020ovm}
Fabian Schmidt.
\newblock {An $n$-th order Lagrangian Forward Model for Large-Scale Structure}.
\newblock {\em JCAP}, 04:033, 2021.

\bibitem{Rampf:2022tpg}
Cornelius Rampf, Sonja~Ornella Schobesberger, and Oliver Hahn.
\newblock {Analytical growth functions for cosmic structures in a
  \ensuremath{\Lambda}CDM Universe}.
\newblock {\em Mon. Not. Roy. Astron. Soc.}, 516(2):2840--2850, 2022.

\bibitem{Garny:2022fsh}
Mathias Garny and Petter Taule.
\newblock {Two-loop power spectrum with full time- and scale-dependence and EFT
  corrections: impact of massive neutrinos and going beyond EdS}.
\newblock {\em JCAP}, 09:054, 2022.

\bibitem{Bryan_1998}
Greg~L. Bryan and Michael~L. Norman.
\newblock Statistical properties of x-ray clusters: Analytic and numerical
  comparisons.
\newblock {\em The Astrophysical Journal}, 495(1):80, mar 1998.

\bibitem{Sheth:2000ii}
Ravi~K. Sheth and Antonaldo Diaferio.
\newblock {Peculiar velocities of galaxies and clusters}.
\newblock {\em Mon. Not. Roy. Astron. Soc.}, 322:901, 2001.

\bibitem{Gasser:1983yg}
J.~Gasser and H.~Leutwyler.
\newblock {Chiral Perturbation Theory to One Loop}.
\newblock {\em Annals Phys.}, 158:142, 1984.

\bibitem{Pajer:2013jj}
Enrico Pajer and Matias Zaldarriaga.
\newblock {On the Renormalization of the Effective Field Theory of Large Scale
  Structures}.
\newblock {\em JCAP}, 08:037, 2013.

\bibitem{Wilson:1974sk}
Kenneth~G. Wilson.
\newblock {Confinement of Quarks}.
\newblock {\em Phys. Rev. D}, 10:2445--2459, 1974.

\bibitem{McNeile:2006qy}
C.~McNeile and Christopher Michael.
\newblock {The Decay constant of the first excited pion from lattice QCD}.
\newblock {\em Phys. Lett. B}, 642:244--247, 2006.

\bibitem{Fukugita:1992np}
M.~Fukugita, N.~Ishizuka, H.~Mino, M.~Okawa, and A.~Ukawa.
\newblock {Pion decay constant in full lattice QCD}.
\newblock {\em Phys. Lett. B}, 301:224--230, 1993.

\bibitem{Mastropas:2014fsa}
Ekaterina~V. Mastropas and David~G. Richards.
\newblock {Decay constants of the pion and its excitations on the lattice}.
\newblock {\em Phys. Rev. D}, 90(1):014511, 2014.

\bibitem{Nascimento:2024rbv}
Caio Nascimento, Drew Jamieson, Matthew McQuinn, and Marilena Loverde.
\newblock {A semi-analytic estimate for the effective sound speed counterterm
  in the EFTofLSS}.
\newblock {\em JCAP}, 02:023, 2025.

\bibitem{peebles1980large}
P.J.E. Peebles.
\newblock {\em The Large-scale Structure of the Universe}.
\newblock Princeton Series in Physics. Princeton University Press, 1980.

\bibitem{Peebles:2009hw}
P.~J.~E. Peebles.
\newblock {Phenomenology of the Invisible Universe}.
\newblock {\em AIP Conf. Proc.}, 1241(1):175--182, 2010.

\bibitem{f2381f32-aadb-3c8a-a473-53334cac15e7}
James Binney and Scott Tremaine.
\newblock {\em Galactic Dynamics: Second Edition}.
\newblock Princeton University Press, rev - revised, 2 edition, 2008.

\bibitem{Philcox:2021kcw}
Oliver H.~E. Philcox and Mikhail~M. Ivanov.
\newblock {BOSS DR12 full-shape cosmology: \ensuremath{\Lambda}CDM constraints
  from the large-scale galaxy power spectrum and bispectrum monopole}.
\newblock {\em Phys. Rev. D}, 105(4):043517, 2022.

\bibitem{Ivanov:2023qzb}
Mikhail~M. Ivanov, Oliver H.~E. Philcox, Giovanni Cabass, Takahiro Nishimichi,
  Marko Simonovi\'c, and Matias Zaldarriaga.
\newblock {Cosmology with the galaxy bispectrum multipoles: Optimal estimation
  and application to BOSS data}.
\newblock {\em Phys. Rev. D}, 107(8):083515, 2023.

\bibitem{Holm:2023laa}
Emil~Brinch Holm, Laura Herold, Th\'eo Simon, Elisa G.~M. Ferreira, Steen
  Hannestad, Vivian Poulin, and Thomas Tram.
\newblock {Bayesian and frequentist investigation of prior effects in EFT of
  LSS analyses of full-shape BOSS and eBOSS data}.
\newblock {\em Phys. Rev. D}, 108(12):123514, 2023.

\bibitem{Simon:2022lde}
Th\'eo Simon, Pierre Zhang, Vivian Poulin, and Tristan~L. Smith.
\newblock {Consistency of effective field theory analyses of the BOSS power
  spectrum}.
\newblock {\em Phys. Rev. D}, 107(12):123530, 2023.

\bibitem{Maus:2024sbb}
M.~Maus et~al.
\newblock {A comparison of effective field theory models of redshift space
  galaxy power spectra for DESI 2024 and future surveys}.
\newblock 4 2024.

\bibitem{Carrilho:2022mon}
Pedro Carrilho, Chiara Moretti, and Alkistis Pourtsidou.
\newblock {Cosmology with the EFTofLSS and BOSS: dark energy constraints and a
  note on priors}.
\newblock {\em JCAP}, 01:028, 2023.

\bibitem{Donald-McCann:2023kpx}
Jamie Donald-McCann, Rafaela Gsponer, Ruiyang Zhao, Kazuya Koyama, and Florian
  Beutler.
\newblock {Analysis of unified galaxy power spectrum multipole measurements}.
\newblock {\em Mon. Not. Roy. Astron. Soc.}, 526(3):3461--3481, 2023.

\bibitem{Mathematica}
Wolfram~Research{,} Inc.
\newblock Mathematica, {V}ersion 14.1.
\newblock Champaign, IL, 2024.

\end{thebibliography}
\bibliographystyle{unsrt}
\end{document}